\documentclass[12pt,preprint]{aastex}

\usepackage{amsmath}
\usepackage{amssymb}
\usepackage{xspace}
\usepackage{color}
\usepackage{ifthen}
\usepackage{xspace}

\newcommand{\kms}{\ensuremath{\mathrm{km~s}^{-1}}}
\newcommand{\gpc}{\ensuremath{\mathrm{g~cm}^{-3}}}
\newcommand{\acc}{\ensuremath{\mathrm{cm~s}^{-2}}}
\newcommand{\Nifs}{\ensuremath{^{56}\mathrm{Ni}}}
\newcommand{\Nife}{\ensuremath{^{58}\mathrm{Ni}}}

\newcommand{\Fefs}{\ensuremath{^{56}\mathrm{Fe}}}
\newcommand{\Feff}{\ensuremath{^{54}\mathrm{Fe}}}

\newcommand{\Msun}{{\ensuremath{\mathrm{M}_{\odot}}}}

\newcommand{\lSect}[1]{{\label{sec:#1}}}
\newcommand{\lFig}[1]{{\label{fig:#1}}}
\newcommand{\lEq}[1]{{\label{eq:#1}}}
\newcommand{\lTab}[1]{{\label{tab:#1}}}

\def\gtaprx {\lower .1ex\hbox{\rlap{\raise .6ex\hbox{\hskip .3ex
	{\ifmmode{\scriptscriptstyle >}\else
		{$\scriptscriptstyle >$}\fi}}}
	\kern -.4ex{\ifmmode{\scriptscriptstyle \sim}\else
		{$\scriptscriptstyle\sim$}\fi}}}
\def\ltaprx {\lower .1ex\hbox{\rlap{\raise .6ex\hbox{\hskip .3ex
	{\ifmmode{\scriptscriptstyle <}\else
		{$\scriptscriptstyle <$}\fi}}}
	\kern -.4ex{\ifmmode{\scriptscriptstyle \sim}\else
		{$\scriptscriptstyle\sim$}\fi}}}
\newcommand{\FIGFF}[2]{{\ref{fig:#2}{#1}}}

\newcommand{\Figure}[1]{{Figure~\FIGFF{}{#1}}}

\newcommand{\TABFF}[1]{{\ref{tab:#1}}}
\newcommand{\Table}[1]{{Table~\TABFF{#1}}}

\newcommand{\Sectff}[1]{{\ref{sec:#1}}}
\newcommand{\Sect}[1]{{\S~\Sectff{#1}}}

\newcommand{\Eqref}[1]{{\ref{eq:#1}}}
\newcommand{\Eqff}[1]{{(\Eqref{#1})}}

\newcommand{\EQ}[1]{{Equation~\Eqff{#1}}}

\newcommand{\figurepath}{./}

\begin{document}

\title{Carbon Deflagration in Type Ia Supernova: I. Centrally Ignited Models} 

\author{H. Ma\altaffilmark{1}, S. E. Woosley\altaffilmark{1},
  C. M. Malone\altaffilmark{1}, 
  A. Almgren\altaffilmark{2}, and J. Bell\altaffilmark{2}}

\altaffiltext{1}{Department of Astronomy and Astrophysics, University
  of California, Santa Cruz, CA 95064} \altaffiltext{2}{Center for
  Computational Sciences and Engineering, Lawrence Berkeley National
  Lab, Berkeley, CA 94720}

\begin{abstract}

A leading model for Type Ia supernovae (SNe Ia) begins with a white
dwarf near the Chandrasekhar mass that ignites a degenerate
thermonuclear runaway close to its center and explodes. In a series of
papers, we shall explore the consequences of ignition at several
locations within such dwarfs. Here we assume central ignition, which
has been explored before, however, the problem is worth revisiting, if
only to validate those previous studies and to further elucidate the
relevant physics for future work. A perturbed sphere of hot iron ash
with a radius of $\sim$100 km is initialized at the middle of the
star.  The subsequent explosion is followed in several simulations
using a thickened flame model in which the flame speed is either fixed
--- within the range expected from turbulent combustion --- or based
on the local turbulent intensity.  Global results, including the
explosion energy and bulk nucleosynthesis (e.g. $\Nifs$ of 0.48--0.56
$\Msun$) turn out to be insensitive to this speed.  In all completed
runs, the energy released by the nuclear burning is adequate to unbind
the star, but not enough to give the energy and brightness of typical
SNe Ia.  As found previously, the chemical stratification observed in
typical events is not reproduced.  These models produce a large amount
of unburned carbon and oxygen in central low velocity regions, which
is inconsistent with spectroscopic observations, and the intermediate
mass elements and iron group elements are strongly mixed during the
explosion.

\end{abstract}


\section{INTRODUCTION}
\lSect{intro}

Common Type Ia supernova (SN Ia) are believed to be the thermonuclear
explosions of carbon-oxygen white dwarfs in binary systems
\citep{Hoy60,Hil00}. Beyond this generality though lies room for
considerable diversity. The white dwarf can grow to criticality by
accreting mass slowly from a companion star, the so called
``single-degenerate'' model, or by merging with another white dwarf,
the ``double degenerate model.'' Depending upon the accretion rate and
angular momentum transfer, the mass of the white dwarf when the
nuclear runaway begins can vary upwards \citep{How06} or downwards
\citep[e.g.][]{Woo11a} from the Chandrasekhar value (1.38 \Msun \ for
ignition).  Ignition itself may happen in the center, slightly
off-center, or near the surface of the white dwarf.  Historically,
greatest attention has been given to the centrally ignited
Chandrasekhar mass model (the MCh model). This focus partially reflected
the predominance, until recently, of one-dimensional (hence
spherically-symmetric) codes and models, but also the success of the
model itself.  Given the liberty of varying uncertain flame physics,
including the possibility of a delayed detonation, the MCh model has
demonstrated striking agreement with the observed energetics, light
curves, spectra, and nucleosynthesis of common SN Ia events
\citep{Nom84,Bra85,Hoe96,Maz07}.

More recently though, it has been realized that other models are
probably needed, especially if one is to explain unusual events
\citep[e.g.][]{How06,Str06,Per10,Pak11}. Simulations with increased
complexity and realism have also shown that the MCh model itself
probably doesn't ignite exactly in its center
\citep{Gar95,Hoe02,Kuh06,Zin11,Non12}.  This symmetry breaking may
have important observational consequences for the explosion mechanism
\citep{Ple04,Rop07b}, as well as the spectrum and light curve
\citep{Kas09}. Still it remains interesting to explore the traditional
model. Despite x-ray limits on the progenitor system \citep{Gil10},
the single-degenerate model remains viable \citep{Vos08,Hac10}, and
even in double degenerate systems, the mergers of white dwarf pairs
may produce MCh models as well as super-MCh models \citep{Yoo07}.
Studies of off-center ignition also show a small, but non-negligible
probability of ignition occurring so close to the center that it
probably burns through to the other side \citep{Zin07,Non12}.

In the last decade, the symmetric MCh model has also attracted
considerable attention from the three-dimensional simulators. Despite
some initial optimism \citep{Rei02,Rop06}, these calculations
\citep{Gam03,Gar05,Sch06,Rop07a} have shown that even though pure
deflagration, i.e., deflagration without a subsequent detonation,
released enough energy to unbind the star, it is unable to provide the
larger kinetic energies characteristic of common SN Ia. Equally
problematic, the deflagration-only models produce too little
$^{56}$Ni, leave behind significant unburned carbon and oxygen with
low speed near the center of the ejecta, and lack intermediate mass
elements at high velocity. These aspects make it difficult for pure
deflagration alone to produce a common SN Ia consistent with
observations, although see \cite{Kro13} for pure deflagration models
of sub-luminous SNe Ia.

Here we revisit this classic problem of centrally ignited deflagration
in MCh models, not with the intention to prove that they work, but to
understand better why they don't work. Our study has several
motivations. First, different groups have assumed different initial
conditions, codes, grid geometries, and resolutions. With the power of
adaptive mesh refinement (AMR) and larger computers, we are able to
define the initial conditions more precisely and track the burning
more realistically. Previous studies also disagree concerning how much
carbon is left at the center after the explosion is over and on the
amount of intermediate mass elements (IME) produced. We want to
understand this and give an alternate set of results. We are also
implementing a novel, inexpensive approach to flame propagation that
uses tables of yields generated off-line assuming isobaric burning
and want to test the sensitivity of
results to uncertain parameters in a well-studied case. Finally, in
order to study delayed detonation in the MCh model as we plan to do in
a subsequent paper, it is necessary to better understand the
conditions that exist at the transition.

In \Sect{code} we briefly describe the fully compressible
hydrodynamics code, CASTRO used in these studies \citep[see
  also][]{Alm10}, and describe our treatment of the flame
physics. Details of the calculations are presented in
\Sect{calculations}, followed by the results of our simulations in
\Sect{results}. Discussion and comparison to previous work are
presented in \Sect{dis}.

\section{The CASTRO Code and Numerical Implementation of the Problem}
\lSect{code}
\subsection{CASTRO}

Because numerous instabilities are involved in the propagation of the
burning and the important role played by turbulence, realistic studies
must be carried out in three dimensions. To avoid any grid
singularities, it is also helpful if the mesh is Cartesian. Coupled
with the requirement that the burning regions themselves be well
resolved implies that this is a supercomputer problem. Even with
today's largest machines, a calculation that resolves the real flame
surface ($\sim10^{-2}$ cm) and covers the full star ($\sim 10^8$ cm)
over the duration of the supernova explosion is impossible.  Instead
we rely on implicit large eddy simulations with adaptive mesh
refinement (AMR) to provide the highest (feasible) resolution only in
the regions where it is most needed.

Over the last five years, our group has developed a code that
specifically fits these requirements. CASTRO \citep{Alm10} is a fully
compressible, finite-volume radiation-hydrodynamics code that uses an
unsplit version of the piecewise parabolic method (PPM). The grid is
Eulerian and uses a hierarchical block-structured approach to
AMR. CASTRO does not require a strict parent-child relation between
patches in different levels (Figure 1 in \citet{Alm10}), and performs
sub-cycling in time-step advance for the refined region. Both features
enhance its efficiency on massively parallel machines.

Self-gravity is supported in CASTRO using a Poisson solver, as well as
an economic monopole approximation, which creates a 1D radial gravity
field consistent with the mass distribution. The latter was used
here. Users may supply external modules to define a suitable equation
of state (EOS) and nuclear reaction networks.  The user can control
the regions that are refined based upon combinations of many physical
parameters, or their gradients. The mesh spacing can vary by a factor
of 2 or 4 between levels. Time steps at each level are constrained by
a standard CFL condition based on maximum wave speed over the grids at
that level.  Additional limitations can be imposed by the user to
follow the temporal evolution of composition changes by nuclear
reactions. Another useful feature is the ability to enlarge the
computational domain from a restart file. In our simulation, the final
domain is 32 times larger than the original domain. Thus while the
grid is not co-moving with the matter, it can follow explosions over a
large increase in the initial radius to obtain a homologous coasting
configuration. Modules can also be provided to describe the initial
model and flame propagation.

\subsection{Mapping and Initialization}

The presupernova model employed here is a 1.38 $\Msun$ white dwarf
with an initial composition of $50\%$ carbon and $50\%$ oxygen,
generated using KEPLER, a 1D implicit Lagrangian code
\citep{Wea78,Woo02}.  This white dwarf model has been previously used
many times as a starting point for studies of SN Ia
\citep[e.g.][]{Nie97,Woo07a} and is similar to the initial model of
Nomoto's Model ``W7'' \citep{Nom84}. The central density of the star
at the time of the final runaway was $2.6 \times 10^{9}$ $\gpc$, the
central temperature, $6 \times 10^8 $ K, and the initial radius, about
1700 km.  To ensure hydrostatic equilibrium on the new Cartesian grid,
the distribution of physical parameters was adjusted from
their values in KEPLER to satisfy the differencing formulae, $0.5
(\rho_i+ \rho_{i-1}) g = (P_{i} - P_{i-1})/dr$. Constant entropy was
assumed in the convection zone which includes most of the mass of the
presupernova star. Here $g= G M(r)/r^2$ was defined on the interface
between zones, where $M(r)$ is the enclosed mass at the radius of $r$.
For the multi-dimensional calculations, a very fine grid was first
generated in hydrostatic equilibrium in the 1D version of CASTRO using
spherical coordinates. This grid was then interpolated onto the
Cartesian grid AMR hierarchy to give the
starting model in 3D.  The
Helmholtz stellar equation of state (EOS) of Frank Timmes was used in
all of our calculations \citep{Tim00}. Because the Eulerian grid
requires non-zero values in all zones, the white dwarf was surrounded
by an artificial, but inert, low-density medium with temperature
$5\times 10^6$ k and density $10^{-4}\ \gpc$. This low density medium
had no dynamic effect on the supernova explosion calculation. A
numerical ``sponge'' was implemented to damp the velocity in the
surrounding medium where densities were between $10^{-4}$ and
$1000\ \gpc$. This was done by multiplying the three velocity
components by a factor that varied smoothly from nearly 0 to 1 in this
density range \citep{Zin09}. The sole purpose of the sponge was to
keep the time step in the low density, artificial circumstellar medium
from affecting the calculation as would happen if the medium began to
accrete onto the white dwarf.

Burning was initiated by a nearly spherical bubble of hot iron ash
surrounding the center of the star with a radius of 100 km. The
temperature in the ash was set to 8.5 billion K, and binding energy
per nucleon, $BE$, and mean atomic weight, $\bar{A}$, were set using
the nuclear burning tables (Tables 1 and 2) to 8.17 MeV/nucleon and
11.6 respectively. The ash density was derived from the equation of
state using this temperature and the same pressure as the surrounding
fuel. Since the initial ash was in nuclear statistical equilibrium
(NSE), its abundances and $Y_e$ started to evolve immediately.  The
surface of this sphere was perturbed using spherical harmonic
functions assuming a superposition of all $l$-wave modes from 14 to 20
and a maximum perturbation amplitude of about 25 km. The choice of 100
km for the radius was influenced by studies of ignition
\citep[e.g.][]{Non12} which showed that prompt ignition outside this
radius was very unlikely. This is a smaller initial radius than has
been employed by some groups \citep{Rop06,Rop07a}. The initial zoning
and AMR structure employed are described in \Sect{mesh}. Pressure was
kept unchanged across the ash fuel interface, though the ash had lower
density and higher temperature. The initial perturbed sphere was
therefore Rayleigh-Taylor unstable, though it took some time for these
initial instabilities to grow. This use of a single large deformed
ignition region seems more physical to us than ignition at many
numerous distinct points \citep{Rop06}, especially since recent
studies show that the flame only ignites once \citep{Non12}.

\subsection{Nuclear Physics of the Burning}

The nuclear physics of high temperature carbon deflagration can be
quite complex. The products of the burning depend upon the density and
may or may not be in NSE. After the deflagration front has passed, the
composition continues to evolve through electron capture and the
reassembly of partially photodisintegrated matter. At low density,
nuclear burning can continue. Electron capture changes the electron
mole number, $Y_e$, to which the pressure is quite
sensitive. Reassembly of alpha-particles into iron-group elements
provides a substantial energy boost as the composition cools by
expansion. Nuclear fusion can appreciably alter the final composition,
especially at lower temperatures where NSE is not achieved. These
effects can only be followed accurately with a a large nuclear
reaction network. Yet at the same time, coupling a large network
directly to a 3D simulation would be very computationally
expensive. Worse still, the nuclear time step can be very small,
especially in NSE, and the solution to the stiff system of ODEs is
prone to numerical instability.

Our solution was to use tables to describe the energy generation and
composition changes. The tables were generated off-line using a large
nuclear reaction network whose accuracy has been tested many times
in the Kepler stellar evolution code. Different tables were used to
describe the burning in the flame \Sect{burning} and the volumetric
burning in NSE \Sect{nse}. The implementation of this burning in a
flame model is discussed in \Sect{flame_model}.

\subsubsection{Burning in the Flame}
\lSect{burning}

The first stage of the burning occurs as the flame passes through the
unburned carbon and oxygen. At the high density of the deep white
dwarf interior, this flame is very thin, though highly convoluted,
essentially a moving unresolved discontinuity that turns carbon and
oxygen into iron-group elements, alpha-particles, and nucleons in
NSE. At lower densities, the heat capacity of the matter rises and the
burning results in a lower temperature. Carbon, oxygen, neon and
silicon burning regions in the flame grow thicker and eventually
burning no longer culminates in NSE, but in the production of
intermediate mass elements, silicon through calcium. At still lower
density, the oxygen does not burn to completion and one has neon,
oxygen and magnesium. Eventually the burning goes out.

A characteristic shared at all densities though is that the burning in
the flame transpires at constant pressure. Detonation does not occur
in the models being studied, and because the flame speed during the
deflagration stage is such a small fraction of the sound speed
($\sim1$\%), pressure equilibrates across the flame. Since the gas is
initially very degenerate and has $Y_e$ = 0.50, the pressure in the
fuel, and hence in the ash, is completely determined by the fuel's
density and is not very sensitive to its (low) temperature. Assuming a
reasonable limit to the time scale for the burning to continue once
the energy generation saturates at a maximum, the change in
composition at a given density can be uniquely determined from the
initial composition and the isobaric condition. This allows the energy
yield and ash composition to be computed off-line using a reaction
network of arbitrary size.

Beginning at a low temperature, taken here to be $6 \times 10^8$ K,
and for a grid of initial densities from 10$^{6.4}$ to 10$^{10.0}$ g
cm$^{-3}$ in steps of log $\rho$ of 0.05, the results of nuclear
burning were calculated using a nuclear reaction network that
contained 188 isotopes, including neutron, proton, $^{4}$He, $^{12-14}$C, $^{14-15}$N,
$^{16-18}$O, $^{18-19}$F, $^{19-22}$Ne, $^{22-23}$Na, $^{23-28}$Mg,
$^{26-29}$Al, $^{27-32}$Si, $^{31- 36}$P, $^{30-33}$S, $^{34-39}$Cl,
$^{35-42}$Ar, $^{38-42}$K, $^{39-48}$Ca, $^{41-49}$Sc, $^{43-52}$Ti,
$^{45-53}$V, $^{47-56}$Cr, $^{49-57}$Mn, $^{51-60}$Fe, $^{53-61}$Co,
$^{55-64}$Ni, $^{57-65}$Cu, $^{59-68}$Zn, $^{61-69}$Ga and
$^{63-72}$Ge.  The initial composition was 50\% $^{12}$C and 50\%
$^{16}$O by mass. Burning proceeded with an imposed isobaric
condition. That is, as time passed the temperature rose and density
fell in response to the energy generation, but the changes in each
were such as to keep the pressure constant for the appropriate
equation of state. The ``Helmholtz'' equation of state was used to
obtain the pressure, its derivatives, and the heat capacity at
constant pressure during this iteration \citep{Tim00}.

For each point in the $\rho-T$ grid, after a lengthy time, which
would be much shorter in the star due to conduction, the temperature
gradually began to rise and the rise accelerated along with the
nuclear energy generation. For each density, there came a point where
the nuclear energy generation reached a maximum and then began to
decline due to the depletion of the most incendiary parts of the
composition. Once the energy generation had declined to 80\% of that
maximum value, a counter was set and 0.1 s later the calculation was
halted and the characteristics of the ash sampled. The 0.1 s is
arbitrary, but the composition and temperature were not changing much
during this time. The value chosen is a lower bound to the
hydrodynamic expansion time of the star and an upper bound to the time
for burning to cross the typical five zone thickness of the
(thickened) flame.

This procedure is an approximation to what would actually happen in a
multi-zone calculation of a steady state laminar flame. In a flame,
conduction, not gradual nuclear burning, would heat the material to
the point that it flashed on a time scale sufficiently short to
sustain a self-similar front with a thickness given by the Landau
condition \citep{Lan59}, which says that the nuclear time balances the
energy diffusion time. Later in the burning, a small amount of heat
would diffuse {\sl out} of the ash into the fuel in order to ignite it
and keep the flame going. However, because of the extreme temperature
sensitivity of the burning rates and the small heat capacity of the
degenerate matter, the amount of heat that must diffuse to propagate
the flame is small. The major part of the burning thus occurs in
thermal isolation at constant pressure and can be approximated by a
single zone as we have done here. It is necessary only to resolve the
energy generation spike that is most sensitive to the density of the
fuel, and the small amount of burning that occurs after that spike. If
the flame were very much broader or slower moving and the star did not
expand so that the 0.1 s time scale was orders of magnitude longer,
the ashes would be different, but the necessary change in time scale
to have a major effect is very large. In the future, if flames need to
be studied in very different circumstances, it would not be difficult
to couple the off-line reaction network to a multi-zone (1D) flame model
\citep{Woo09}.

The validity of the approximation is demonstrated by the good
agreement with previous fine-zoned studies of laminar flames under the
same conditions. \Table{compare} shows the comparison with other
studies that also used either a 50\% C, 50\% O composition
\citep{Tim92,Woo07} or a 40\% C, 60\% O composition \citep[][indicated
  with an asterisk]{Woo11b}. Overall the agreement between ash temperature and
density changes found in these studies is excellent. For the limited
compositional information given in these references, the agreement was
also very good.

The resulting composition from our network simulations was packed into
the seven abundance variables carried by the 3D code depending on the
atomic weight of the product nucleus.  Species lighter than A = 12
were packed into helium; species from 12 through 15 into carbon; 16
through 19 into oxygen; 20 through 23 into neon; 24 through 27 into
magnesium; and 28 through 47 into silicon. Hence all intermediate mass
elements, silicon through calcium are contained in the ``silicon''
variable. Species heavier than A = 47 were generically considered to
be ``iron'' (\Table{models1}). Additionally the final temperature,
density, binding energy per nucleon, and mean atomic weight, $\bar A$,
were determined for the ash (\Table{models2}).

Once the change in energy and composition across the flame is known as
a function of the initial density, the burning can be implemented
without any recourse to temperature-dependent reaction rates or
network solution. The flame is essentially a thin phase transition moving
with a desired speed. The interface can be treated either using a
level-set approach, or as was done here, as a thickened flame. The
progress variable is the carbon mass fraction. As its mass fraction
varies from 0.5 to zero, the binding energy per nucleon changes
linearly from its value in the fuel, 7.828 MeV per nucleon, to its
value given for carbon--free ash in the tables. The seven abundance
variables are interpolated linearly from their values in the
fuel (zero for all but C and O) and in the ash. The nuclear energy
generation rate was

\begin{equation}
\label{enuc}
\epsilon_{\rm nuc} = 9.64 \times 10^{17} (\Delta BE/0.5)(d X(^{12}C)/dt) \ \ {\rm
  erg \ g^{-1} \ s^{-1}}.
  \end{equation}

Here 0.5 is the initial mass fraction of carbon and $\Delta BE$ is the
change in binding energy per nucleon following the burning of a
composition of 50\% C and 50\% O to ash (\Table{models2}).  Though
the oxygen usually burns as well, carbon was the progress variable. The
change in the carbon abundance itself was calculated using a thickened
flame model (\Sect{flame_model}).  Several one dimensional studies
were carried out with the CASTRO code that showed this prescription
for energy generation and composition change gave the same resultant
temperatures and densities for a conductive flame as in the one zone
network simulation.

For fuel densities below 10$^{6.5}$ g cm$^{-3}$ it was assumed that no
nuclear burning took place and no burning was allowed on the grid
until conduction had raised the local temperature in a carbon-rich
composition above $2 \times 10^9$ K.

\subsubsection{Volumetric Burning in Nuclear Statistical Equilibrium}
\lSect{nse}

In NSE, all strong and electromagnetic nuclear reactions have come 
into a state of near balance with
their reverse reactions, e.g. (n,$\gamma$), (p,$\gamma$), and
($\alpha,\gamma$) reactions are balanced by ($\gamma$,n),
($\gamma$,p), and ($\gamma,\alpha$) reactions. The distribution of
abundances is then completely specified by just three parameters: the
temperature, density and the neutron excess. The latter is usually
measured by the electron mole number,
\begin{equation}
Y_e= \Sigma Z_i Y_i
\end{equation}
where $Y_i$ is the mass fraction of the species ``i'' divided by its
atomic weight and $Z_i$ is its nuclear charge. For the initial
composition and for the ash immediately behind the flame, there is
neutron-proton equality so that $Y_e$ = 0.50.

The NSE composition can evolve, however, both as a consequence of the
changing temperature and density and the decrease of $Y_e$ due to
electron capture on nuclei at high densities $\rho \gtaprx 10^8$ g
cm$^{-3}$. Of particular importance is the fact that, at the high
temperatures appropriate to the central regions of exploding white
dwarfs, NSE favors the partial photodisintegration of iron group
nuclei into alpha-particles. This tends to reduce the initial increase
in nuclear binding energy and decrease the average atomic mass of the
ions. The temperature of the ash is not as high as one would obtain
for fusion just to iron. However, as the matter expands and cools the
alpha-particles reassemble, providing late time energy input to the
explosion. Also important is the reduction of $Y_e$ by electron
capture. Since relativistic electrons are providing the pressure and
since that pressure depends on
the combination $(\rho Y_e)^{4/3}$, this can affect the dynamics. At
constant pressure, a decrease in $Y_e$ means an increase in the
density and a reduction in buoyancy so plumes rise less rapidly. The
value of $Y_e$ also affects the nucleosynthesis and light curve of the
supernova. For values close to 0.50, the principle nucleosynthetic
product is radioactive $^{56}$Ni, but for $Y_e$ \ltaprx 0.48,
$^{54}$Fe and $^{58}$Ni are more abundant.

The evolution of the NSE composition can only be followed using a
large network of nuclei since many nuclei participate in capturing
electrons at different densities and many reactions are involved in
maintaining the equilibrium. We used the 127 isotope equilibrium
network from the Kepler stellar evolution code \citep{Wea78} with weak
interaction rates taken from a variety of sources as discussed in
\citet{Heg01}. Particularly important are the revisions to earlier
rates by \citet{Lan00} and the inclusion of accurate rates for
beta-decay and positron emission as well as electron
capture. Temperature dependent partition functions were used in
calculating the abundances. The isotopes included in the NSE network
were neutrons, $^{1-3}$H, $^{3-5}$He, $^{5}$Li, $^{9}$Be, $^{12}$C,
$^{16}$O, $^{20}$Ne, $^{23}$Na, $^{24-26}$Mg, $^{26-28}$Al,
$^{28-30}$Si, $^{30- 33}$P, $^{31-34}$S, $^{35-37}$Cl, $^{36-40}$Ar,
$^{39-43}$K, $^{40-48}$Ca, $^{44-49}$Sc, $^{44-52}$Ti, $^{47-54}$V,
$^{48-56}$Cr, $^{51-58}$Mn, $^{52-62}$Fe, $^{54-64}$Co, $^{56-66}$Ni,
$^{59}$Cu, and $^{60}$Zn. 

The table was constructed to include, as a function of the three
independent variables, $\rho$, T, and $Y_e$, the binding energy per
nucleon of the NSE distribution, its average atomic weight, $\bar A$,
the helium mass fraction, and the time rate of change of $Y_e$
resulting from all the weak interactions. Thirty one values of density
from 10$^7$ to 10$^{10}$ g cm$^{-3}$, 71 temperatures from 10$^9$ to
10$^{10.4}$ K, and 21 values of $Y_e$ from 0.50 to 0.40 were
included in the table (46221 grid points). The temperature and density
grid was in equal steps of the base 10 log. The values of $\Delta Y_e$
were also equal so table look up time was minimized. Linear
interpolation in log T, log $\rho$ and $Y_e$ thus gave the binding
energy, mean atomic mass, and time rate of change of $Y_e$ for any
point of interest in the calculation. NSE was only assumed to exist in
a zone if its density was over 10$^8$ g cm$^{-3}$, its temperature
above $3 \times 10^9$ K, its carbon mass fraction was below 1\%, and
the combined mass fractions of iron and helium were over 88\%.

Due to the stiff dependence of abundances and binding energy on
temperature and density, NSE calculations are subject to a well known
instability. If the temperature goes up at constant density, for
example, photodisintegration can give a large negative energy
generation rate. That will tend to reduce the temperature for the next
call which results in recombination and a large positive energy
generation. Since the time scale for this adjustment is instantaneous,
the code may take very small times steps and still have convergence
problems. This difficulty was circumvented here by the addition of
numerical inertia to the NSE calculation. The change in mean binding
energy implied by a temperature-density change was not applied fully
in a single time step but spread over three time steps. In practice,
this kept the calculation stable and asymptotic values for $Y_e$ and
the binding energy were attained much more quickly than the time scale
for the temperature or density to vary appreciably.

A few sample points from the NSE table are given in \Table{nse}. In the
actual table, six figure precision was maintained.  The NSE network
was not large enough, nor reliable weak interaction information
available for the accurate calculation of abundances and $\dot Y_e$
for $Y_e$ below 0.43. Fortunately, the smallest value of $Y_e$ in any
of our simulations here was above 0.45.

\subsection{Flame Model}
\lSect{flame_model}

Burning during the deflagration stage of a SN Ia is quite complex and
has challenged model builders for years. The most interesting burning
during this stage happens at sufficiently low Karlovitz number, $Ka
\ltaprx 1$, that the concept of a well defined laminar flame remains
valid.  At higher values of $Ka$, which happen for densities below
about 10$^7$ g cm$^{-3}$, turbulence may lead to mixing, at the
microscopic level, of cold fuel and hot ash and a possible detonation
\citep{Nie97,Kho97,Woo09}. That possibility is ignored here. Thus the
picture of the flame is of a surface deformed by floatation, various
instabilities, and possibly turbulence. Large scale motions caused by
buoyancy are captured on the grid, but the smaller scale wrinkling is
not. The manner in which this unresolved burning is treated is usually
called the ``subgrid model'' for burning.  Following \citet{Dam40}, it
is often assumed that if the large scale motions governing the spread
of the flame are sufficiently resolved, e.g., the integral scale of
the turbulence in the vicinity of the flame is well resolved, small
scale burning will adapt so as to ``digest'' any entrained material at
the necessary rate.

Adapting tools developed by the chemical combustion community, two
approaches have been used by astrophysicists. One approach uses a
``level set'' \citep{Osh88,Rei99} and a subgrid model for turbulence
\citep{Sch06a,Sch06b}. The ash-fuel interface, found by solving a
``G-equation'', is moved at a speed given by sampling the velocity
field on the grid and estimating the turbulent intensity. Typical
speeds are $\sim$100 km s$^{-1}$, though a wide range of values is
found \citep{Rop07}. An alternate approach is the ``thickened flame
model'' \citep{Dav50,Sha84}, where an artificial conductivity and
energy generation are employed to spread the flame over several zones
and force it to move at a prescribed speed. Both the flame speed and
width can be much greater than the corresponding laminar values. The
width is determined by the need to resolve the flame while the speed
is sometimes taken to be the value to ``burn out'' structure at the
grid level. It is well known \citep[e.g.][]{Tim92}, that flame
perturbations smaller than
\begin{equation}
\lambda_{\rm min} \approx 4 \pi V^2 \left(g \frac{\delta \rho}{\rho}\right)^{-1}
\lEq{lamdamin}
\end{equation}
will be burned before
they can grow. This is sometimes called the ``fire-polishing
length''. Conversely if one sets $\lambda_{\rm min}$ equal to the
resolution, say a few zones, one can solve for a characteristic speed,
V. Thus $V \sim f \sqrt{g_{\rm eff} \Delta x}$ where $g_{\rm eff} = g
(\Delta \rho)/\rho$, $\Delta x$ is the grid resolution, and f $<
1$. This approach has been adopted by most groups studying the problem
in the US \citep{Kho95,Kho96,Gam03,Tow07}. For a resolution of several
km and $g_{\rm eff} \sim 10^9$ cm s$^{-2}$, this prescription also
gives a characteristic speed $\sim 100$ km s$^{-1}$.

While we see great merit in the level set approach and are working on
developing our own, the current study used a thickened flame.
In particular, the nuclear burning time scale and diffusion time scale
are defined here as,
\begin{equation}
\tau_{nuc} = \tau_{diff} = \frac{n\Delta x}{V},
\end{equation}
where $n$ is the desired number of grid zones across the flame surface
and $\Delta x$ is the grid spacing, so that $n\Delta x$ is the desired
flame width. $V$ is the desired flame speed. The nuclear energy
generation rate and the conductivity are then derived from prescribed
flame width and velocity.

In the flame, where the carbon abundance is larger than 1 percent and
temperature is beyond 2 billion degrees, the carbon consumption rate is
defined as,
\begin{equation}
\frac{d X(^{12}C)}{dt} = - \frac{0.5} {\tau_{nuc}},
\end{equation}
where 0.5 is the initial carbon mass fraction. Additionally,
\begin{equation}
\frac{d(BE)}{d X(^{12}C)} = - \frac{\Delta BE}{0.5},
\end{equation}
so that nuclear burning rate in Equation \eqref{enuc} can be derived.

For partially burned cells inside the flame, the current $BE$ and
$\bar{A}$ are interpolated using the equations:
\begin{equation}
\frac{BE - BE_{f}}{X(^{12}C)-0.5} = -\frac{BE_f-BE_i}{0.5},
\end{equation}
\begin{equation}
\frac{\bar{A} - \bar{A}_{f}}{X(^{12}C)-0.5} = -\frac{\bar{A}_f-\bar{A}_i}{0.5}.
\end{equation}

The opacity is also defined using $\tau_{\rm diff}$ from the thickened
flame model as
\begin{equation}
\kappa = \frac{c \tau_{\rm diff}} {3 \rho \Delta^2},
\end{equation}
and conductive coefficient is,
\begin{equation}
D = 4acT^3/3 \rho \kappa,
\end{equation}
where $c$ is light speed, $\Delta$ is the flame thickness, and $a$ =
$7.5646 \times 10^{-15} \rm{\ erg \ cm^{-3} \ K^{-4}}$ is radiation
density constant. The conductivity is only taken to be non-zero in
zones whose temperature is above 1 billion degree.
This criterion allows us to distinguish the flame and post-flame region from
material heated diffusively in front of the flame.  Suppressing spurious
burning in the diffusion front ahead of
burning front is necessary to prevent an artificial ``diamond''
shape from appearing in the propagation of the flame.

From a survey of one dimensional flames using the above prescription,
we determined the multiplier, $f_w$ and $f_v$ (both of order unity),
of the flame width and speed to get prescribed values at different
densities (see \Table{laminar_parameter}). In the calculation, the
left side is set to be hot ashes, and right side is fresh fuel
(\Figure{f1}).  The boundary condition is impermeable at the
right side, free flow at the left side. Flame speeds are measured with
respect to the fuel (\Figure{f2}).

\subsection{Choice of Flame Speed Parameter}

It thus remains only to pick the desired flame speed. As discussed in
\Sect{flame_model}, previous studies using a subgrid model to describe
the turbulent flow \citep{Rei99,Rei02,Rop05} give a speed on the order
of 100 km s$^{-1}$. In particular, \citet{Rop07} studied the
probability density function (PDF) of turbulent velocity at a density
interval of 1--3$\times 10^7$ $\gpc$ in the deflagration phase. His
plot of the PDF shows a turbulent velocity peaking approximately
100--300 $\kms$. Thickened flame models for typical values of
resolution give a comparable, though somewhat smaller value for V. The
laminar speed, which sets a lower bound to the overall flame speed, is
almost this large at the high density where burning begins. At a
density of $2 \times 10^9$ g cm$^{-3}$, the laminar speed is 76 km
s$^{-1}$ \citep{Tim92}, though it declines rapidly for densities much
less than 10$^9$ g cm$^{-3}$. By the time the density has declined
though, floatation is dominating with typical speeds of thousands of
km s$^{-1}$ and the turbulent estimate or the fire polishing estimate
is valid.

This all suggests that a constant flame speed of 100 km s$^{-1}$ is a good
place to begin and this is the value used in our standard
calculation. However, we also carried out other otherwise identical
calculations using fixed V = 50 and 200 km s$^{-1}$ . The lower value here
is actually a little less than the laminar speed for a composition of
50\% carbon, 50\% oxygen at a density of $2.4 \times 10^9$ g cm$^{-3}$
where the laminar speed has a value of 89 km s$^{-1}$
\citep{Tim92}. However, this high density and laminar speed are only
maintained a short time and, in the interest of maintaining a constant
value of V throughout the run, the lower value of 50 km s$^{-1}$ was
used.  Interestingly, had the carbon mass fraction been 20\%, 50
km s$^{-1}$ would have been the correct value even at this high
density.  

To test the sensitivity of the results to the local turbulence, we
also performed a simulation where the flame speed depended quite
simply on the local turbulent intensity.  In particular, for a zone
within the flame (as described by the nuclear burning regime discussed
in \Sect{burning}) the turbulent velocity fluctuations,
$v_\textrm{turb}$, are calculated in the usual way:
\begin{equation}
\label{vturb}
  v_\textrm{turb} = \left(\overline{v^2} - \overline{v}^2\right)^{1/2},
\end{equation}
where the overbar denotes spatial averaging over a cube of $3^3$ zones
centered about the zone in question.  
The value of
$v_\textrm{turb}$ from \eqref{vturb} is then rescaled to the grid
scale assuming Kolmogorov scaling.  The local flame speed, V, is then
set to the maximum between the laminar speed and the local turbulent
intensity.  As we shall see the answer is gratifyingly insensitive to
the choice of V---constant or set to the local turbulent
intensity---within the limited range explored.

\section{Calculations and Analysis}
\lSect{calculations}

\subsection{Calculations}
\lSect{mesh}

The initial computational domain of the 3D calculations is 8192 km in
each dimension with $512^{3}$ cells at base level so that the largest
cells were (16 km)$^3$. We then added 4 levels of refinement using AMR
so that the resolution at the finest level was (1 km)$^3$.  The finest
resolution was always used in the vicinity of the flame. Specifically,
fine zoning was used in the vicinity of any transitions in carbon mass
fraction from 0.50 to less than 0.49.  Higher resolution was also used
in the highest density regions in order to accurately describe the
nuclear evolution and floatation of the first ashes. The coarsest
resolution was used in the near vacuum outside the exploding star.  In
particular, a region was tagged for refinement when $\rho \Delta x^3 >
2 \times 10^{25}$ g, where $\Delta x$ is the cell size. In order to
maintain appropriate resolution as the flame surface area increased
rapidly, the finest zoning was coarsened twice, once when total number
of cells at all levels was more than 1 billion at 0.63 s, and again
when there were 1.5 billion cells at 0.87 s for the standard run. The
zoning at the finest level was never more than 4 km until the
completion of intermediate mass element production when the
supernova's radius was about 10,000 km.

As noted previously, a constant flame speed independent of the grid
scale was assumed for all (hereafter, model A series) but one
simulation (hereafter, model BT) that used the local turbulent
speed. Three different choices were used for constant flame speed in
three independent runs: $50$ (model A50), $100$ (model A100) and $200$
$\kms$ (model A200). The entire domain was remapped twice for models
A50 and A200 to a grid twice as large each time, by adding another
coarser AMR level than the previous base level, when the outer edge of
the supernova expanded close to the grid boundary. For model A100, the
domain was remapped 5 times in order to follow the expansion to the
homologously expanding state.  Thus our biggest domain size was
262,144 km, 32 times larger than the original domain.  For the A50 and
A200 runs, we only followed the explosion for 1.8 s until the nuclear
burning had shut off due to expansion. The simulation for model BT
used slightly different random initial spherical harmonic
perturbations than the A series of models.  To compare model BT with
the constant flame speed runs, we performed an additional simulation
with the same initial conditions as model BT, but with constant
V=$100\ \kms$, hereafter model B100.  That is, the only difference
between models A100 and B100 are the initial random perturbations of
the ash.  Models BT and B100 were only evolved until the first buoyant
plume reached the surface, and thus did not require remapping to a
larger domain.  All calculations were typically run on 6,144 or 12,288
processors for 2--3 Million CPU hours each on Jaguarpf/Titan at the
OLCF and on Franklin and Hopper at NERSC.

\subsection{Parsing Iron}

Once the composition had ceased to evolve, the electron mole number,
$Y_e$, and helium mass fraction were used to distinguish the dominant
forms of iron ($^{54,56}$Fe, $^{56,58}$Ni) produced in the
explosion. Three cases are considered based on neutron excess $\eta =
1-2\ Ye$ \citep{Har85},
\begin{equation}
\eta=\sum \frac{(N_{i}-Z_{i}) \ X_{i}} {A_{i}}.
\end{equation}

For the conditions studied here, most of the neutrons will be bound in
$^{54}$Fe, $^{56}$Fe, and $^{58}$Ni with the relative proportions
depending upon the values of $\eta$ and the helium mass fraction,
X(He). For near proton-neutron equality, i.e., $\eta$ less than 1/27,
and X(He) less than 0.20, it is assumed that most of the neutrons
reside in $^{54}$Fe with a small fraction in $^{58}$Ni. Specifically
$\eta$ = 1/27 X($\Feff$) + 1/29 X($\Nife$) and X($\Nife$) = f
X($\Feff$) with f = 0.35
\citep{Har85}. If X(He) is larger than 20 percent, however, the
so-called $\alpha$-rich freeze out is assumed to produce mostly
$^{58}$Ni so that f = 10.  For $\eta$ less than 1/14, but greater than
1/27, X($\Nifs$) = 0 and the sum of the $^{54}$Fe, $^{56}$Fe, and
$^{58}$Ni mass fractions is 1 -X(He). Neutron conservation additionally
gives $\eta$ = 1/27 X($\Feff$)+1/29 X($\Nife$)+1/14 X($\Fefs$), and
the condition X($\Nife$) = f X($\Feff$), as before, can be solved to
give the 3 isotopic abundances separately. For $\eta$ greater than
1/14, which was only true for a small fraction of the matter where
burning happened at very high density, $\Fefs$ was assumed to be the
only isotope for iron group elements.

\section{Results}
\lSect{results}

\Figure{f3} shows the propagation of the flame front at different star
times for our standard run, model A100. The flame is wrinkled and
distorted by the Rayleigh-Taylor instability, with buoyant hot ash
bubbles floating toward the surface and spreading while cold, denser
matter moves outwards more slowly and is overtaken.  Viewed in
Lagrangian coordinates, the unburned matter appears to migrate
inwards. At very early times, one sees finger-like structures
resulting from the initial perturbations, but these quickly mushroom
and become non-linear. New plumes form and the structure becomes
complex.

During the early growth of the reactive RT instability, burning
effectively smooths out perturbations below the fire polishing length
(\EQ{lamdamin}).  As the RT modes begin to merge, interact, and break
down, they become wrinkled on ever smaller scales \citep[e.g.]{Zin05}.
Indeed, the flame experiences the effects of turbulence on scales down
to the Gibson scale---the scale at which the laminar speed equals the
turbulent speed.  Furthermore, the turbulent scaling study of
\citet{Cia09} shows that the nature of the turbulence shifts from
being dominated by RT scaling (\EQ{lamdamin}) to Kolmogorov scaling
for isotropic eddies on scales smaller than $\sim15$ km, or about the
initial $\lambda_{\rm min}$.  It should be noted that with our choice
of grid resolution in the present studies, we are not resolving the
turbulent flame on scales smaller than $\lambda_{\rm min}$.

Several effects lead to the evolution of $\lambda_{\rm min}$. As the
flame moves out, the greater interior mass enclosed tends to increase
the gravitational acceleration, but at the same time the star expands,
causing it to decrease. The density contrast, $\Delta \rho/\rho$ also
gets larger at lower density. The combined effect is to decrease
$g_{\rm eff}$ slowly, though its value remains roughly constant at
about 10$^9$ cm s$^{-2}$ throughout much of the burning. Since the
flame speed is also a constant, 100 km s$^{-1}$ in this case,
$\lambda_{\rm min}$ does not vary greatly from about 10 km. While
there are always at least several zones within $\lambda_{min}$, this
does not necessarily mean that structures of this scale are well
resolved. Since the thickened flame itself is typically a few to five
zones wide and any structure must be bounded on all sides where there
is fuel by flames, $\lambda_{\rm min}$ is only marginally
resolved. The smallest structures in \Figure{f6} are roughly 40 km
across. The length scale $\lambda_{\rm min}$ is more fully resolved
for the model A200 run, but less resolved for model A50. Ideally,
greater resolution should be employed, but it would be prohibitively
expensive to run with even twice the current resolution since run time
scales as the fourth power of the resolution, three for the three
dimensions plus one for the shorter Courant time step. Fortunately,
the most interesting burning happens on the finest grid scale, which
is 1 to 2 km during the nucleosynthesis epoch. By 1.2 s when the
finest grid becomes 4 km, $g \sim 8 \times 10^{8} \acc$ and $\Delta
\rho/\rho \sim 0.3$, so $\lambda_{\rm min}$ has already increased to
50 km. Visually, one also sees more structure in the model A50 run
(\Figure{f6}), suggesting that the structure is sensitive to V and we
are actually resolving $\lambda_{\rm min}$.

The flame first breaks out of the surface in the model A100 run 1.21 s
after ignition, by which time the star has already expanded by a
factor of about 2.5 and the central density has declined to $5 \times
10^{7}\ \gpc$.  This decrease in density significantly affects the
products of the burning and allows the production of significant
intermediate mass elements. The largest speed up to this point is
$\sim$ 10,000 $\kms$ in the outermost layers of the star, but the flow is still
subsonic in most of the star. Nuclear burning is over at this point
and, as the internal energy converts into kinetic energy, the maximum
speed rises to 25,000 km s$^{-1}$ at 2.76 s. 

\Figure{f4} shows how kinetic energy, binding energy and internal
energy evolve with time. At the beginning of the explosion, the net
binding energy, i.e., the gravitational binding energy minus the
internal energy, is about $5 \times 10^{50}$ ergs. The final kinetic
energy at infinity is about $6 \times 10^{50}$ ergs, so the total energy
released thermonuclear burning energy is $1.1 \times 10^{51}$
ergs. The supernova ejecta is mainly composed of 0.63 $\Msun$ iron
group elements and 0.16 $\Msun$ IME.  The mass of $\Nifs$ and the
final kinetic energy suggests a relatively weak event in SN Ia.

A sequence of plots in \Figure{f5} shows the evolution of the
angle-averaged supernova composition. At early time, iron group
elements are the major products, but after the flame spreads to the
low density region (less than $5\times10^7\ \gpc$) at 1 s, more IME
are made.  As the iron-group ashes continue to move out at higher
speeds than the unburned fuel, boosted along by the recombination of
helium, there is a gradual accumulation of carbon and oxygen near the
center of the explosion. At the end, about 20\% unburned carbon and
oxygen are left in the central region.

The calculation was repeated using two different flame speed
parameters, models A50 and A200, and followed until 1.8 s when
the nuclear burning was over (but not into the final coasting
phase). Comparisons of the flame front morphology using the different
flame speeds are shown in \Figure{f6}. Slower flame speeds result
in increased structure as noted, but as \Figure{f7} shows, the
the total amount of burned material is roughly the same for flame
speeds of 100 $\kms$ and 200 $\kms$. The slower burning speed is
compensated for by a larger surface area. For a flame speed of 50
$\kms$, the calculation produces 0.73 $\Msun$ iron, 0.17 $\Msun$ IME,
and released $1.3\times10^{51}$ ergs. This is significantly more
burning than either of the other two series A runs but still within 10\% of
their average. We do not know if this limited variation is due to the
small number of runs or reflects real physical dependence on the
parameter. There is a tendency, however, for more burning to occur in an 
explosion that starts slower because it remains denser longer.

The B series simulations used a different random number generator seed
than the A series simulations; the random number generator applied
spherical perturbations of random size and amplitude to the initial
flame surface.  \Figure{f8} compares the $X(^{12}\rm C)=0.49$
isosurface of the constant flame speed model B100 (left) to that of
the turbulent flame speed model BT (right) at 0.3 s (top), 0.5 s
(middle), and 0.7s (bottom).  Morphologically, models B100 and BT are
nearly identical.  A comparison of the left column of \Figure{f8}
(B100) with the center column of \Figure{f6} (A100) at the same time
elucidates the difference in initial model conditions between the two
series. The A series perturbations are all nearly of the same
magnitude, whereas the B series initially has a few slightly dominant
perturbations that grow more rapidly and develop into large-scale RT
plumes; this feature is similar to published 2d studies of central
ignition with perturbed spheres \citep[for example]{Jac10,Kru12}.  The
extraordinarily large appearance of the plume in the lower edge of the
plots is a result of that plume's coming towards the viewer.
Neglecting the dominant plumes, at any given time the size of the ash
bubble is about the same in all runs.

\Figure{f9} shows the total mass of iron group elements (red) and IME
(blue) produced for the B series of models.  The dotted lines are for
model B100, and the dashed lines are model BT.  The two thin, solid
lines show the percent difference between the two models where
positive (negative) values indicate model B100 producing more (less)
material than model BT.  These two different flame models agree
remarkably well.  Initially, the constant flame speed for model B100
is greater than the laminar flame speed, which is the speed at which
model BT burns in the absence of turbulence.  This difference causes
model B100 to burn more material more rapidly than model BT.  By about
0.2 s, some zones within the flame experience turbulent velocities
that are greater than the local laminar flame speed and/or the
constant 100 $\kms$ flame speed of model B100; the flame in model BT
begins to increase in speed, which burns more material that then
decreases the difference between the two flame prescriptions.  As the
flame grew in size, the number of fine-resolution zones increased.
After $\sim0.31$ s of evolution, there were over one billion zones in
the finest level and we de-refined the simulation for computational
efficiency.  This increase by a factor of two in zone size (in each
direction) caused a corresponding increase in the turbulent intensity
on the grid scale by a factor of $2^{1/3}$ according to the Kolmogorov
scaling we assumed in our turbulent flame speed model.  This is
evident in \Figure{f9} as a small but sudden increase in the percent
difference between elemental yields with the constant flame speed
model and the turbulent flame speed model.  Another de-refinement
occurred around 0.61 s, but the change in percent difference between
the models is lost in the noise.

Around 0.4 s, the turbulent flame in model BT has ``caught up'' to the
constant flame speed model B100 in terms of total mass produced in
each elemental group.  The probability density functions for the
turbulent speed at various times in \Figure{f10} show that indeed the
most likely turbulent flame speed used around this time in model BT is
$\sim100 \kms$ as is used in model B100.  The distribution does
have an extended tail, which shows some subset of the zones within the
flame experience turbulent intensities greater than the constant flame
speed model B100.  This causes the turbulent flame speed model BT to
overproduce iron and IME relative to model B100 (negative percent
differences in \Figure{f9}).  It is around this point in the evolution
that simple interpretation of the differences in iron and IME yields
between the two flame speed prescriptions becomes difficult.  There is
a constant battle between 1) the small differences in flame structure
being amplified by the (dominant) acceleration due to buoyancy, 2) the
dominant plumes reaching a region where the density gradient becomes
steep, and 3) the shear induced turbulence that tends to redistribute
the turbulent flame speed to ever  higher velocities.  Even
though the total mass of iron and IME is increasing in both models,
the differences between model B100 and BT fluctuate between positive
and negative due to the interplay of the above mentioned forces.

Throughout the B series runs, the choice of flame speed model
(constant vs. turbulent) made less than a 7\% difference in the iron
and IME group total yields.  Comparing, however, slightly different
initial conditions but the same (constant) flame speed model---models
A100 (\Figure{f7}) and B100 (\Figure{f10})---at, say, 0.6 s shows
nearly a 50\% difference in total iron group yield.  This seems to
suggest that the simulation results are much more sensitive to the
initial conditions of the hot ash than they are to the flame speed
prescription, at least in the case of a constant flame speed model and
a model where the flame speed depends simply on the turbulent
intensity on the grid within the flame.

\section{Discussion and Conclusions}
\lSect{dis}

The elemental yields of our models and the nuclear energy released by
each are given in \Table{products}. It is important to note that the B
series models were not run to completion so their tabulated elemental
yields should not be compared with the other models in the table, but
only with each other to assess the role of a simplified turbulent
flame model.  Results from the A series studies with V = 100 and 200
$\kms$ resemble each other closely and are also similar to those of
\citet{Rop06}. They do show, however, significantly less production of
intermediate mass elements than \citet{Rop07a}. We do not know the
specific cause of the large production of intermediate mass elements
in the R\"opke study, but note that they are significantly larger than
in any other study. Perhaps this reflects the initial conditions and
igniting a very large number of ignition points or an unusually high
flame speed from the turbulent subgrid model at low densities (see
below).  \citet{Gar05} found around 0.45 $\Msun \Nifs$ was produced in
the explosion with $\sim 15$ initial igniting bubbles near the center,
which is similar to our series A results, but IME production in their
simulation was much smaller. \citet{Gam05} found only 0.47 $\Msun$
iron group elements were produced during the deflagration assuming an
initial ignition region with a radius of only 30 km.

The actual mass of $\Nifs$ implied by observations of SN Ia is highly
variable ranging from 0.1 to 0.9 \Msun \citep{Maz07}, but a typical
value is about 0.6 $\Msun$ \citep{Bra95}. A lower limit to the total
burned mass derived from observations of the velocity of silicon,
applying the density-velocity structure in W7 model, is around 1.05
\Msun \citep{Maz07}. Our series A calculations found that
approximately 0.9 - 1.0 $\Msun$ of the 1.38 \Msun \ white dwarf was
burned to heavier elements regardless of the assumed flame speed, not
too much smaller than the optimal value.  This burning certainly
released enough energy to unbind the star, but is well short of what
is needed to account for the brighter SN Ia.

Perhaps this is not too surprising. If typical SN Ia ignite off center
and later detonate, then one expects a correlation between the
displacement of the ignition and the brightness of the
supernova. Explosions igniting off-center burn less before detonating
and are thus denser when the detonation occurs. They thus make more
$^{56}$Ni and are brighter. Central ignition is a limiting case where
the burning in the deflagration stage is maximal. A subsequent
detonation makes less $^{56}$Ni and the total synthesis may be less,
hence a fainter supernova. The puzzle then is not why deflagrations
are faint, but the origin of some SN Ia that are fainter still.

The main difficulty faced by pure deflagrations though is not their
brightness, but their spectrum. They have too much low velocity
unburned fuel near their middles \citep{Mar03}.  More than 70\% of the
central ejecta in our standard model A100 was carbon and oxygen.
Even though the central regions initially burned to completion, unburned fuel
is later found near the center of the star because
the ashes have floated above the fuel, leaving it behind
as they move outward with greater speed than their cold surroundings
(see \Figure{f11}). Iron group elements and IME in the ejecta
continue mixing radially even after burning has gone out; the layered
structure of different chemical products present at the end of burning
but before much expansion is hardly seen. This is contrary to the
observed spectral lines of carbon and silicon in the supernova ejecta,
which indicates velocities mostly beyond 10,000 $\kms$
\citep{Fil97,Bra03,Ben05,Par11}. Previous numerical studies have
demonstrated the same shortcomings out of the centrally-ignited pure
deflagration model. For example, \citet{Gam03} assumed a centrally burned
region of radius 30 km, and carried out the following deflagration
simulation with grid-length and effective gravity-dependent flame
speeds in the thick-flame model. 80\% to 90\% percent of the low
velocity materials near the center are composed of carbon and oxygen
at 1.9 s in their study. Other numerical works initialized
a number of hot bubbles around the center instead of the whole
central region, including 2D and 3D studies using a different flame
tracking method and subgrid model \citep{Rei02,Rop06}, and 3D
calculations using a smoothed particle hydrodynamics (SPH) code
\citep{Gar05}. Central carbon concentrations were found in all these
studies. However, parameter studies have shown that larger numbers
(more than 150) of initially Gaussian distributed hot bubbles with a
maximum ignition radius of 180 km leave much less unburned fuel in
central regions, resulting a spectrum better consistent with the
observations \citep{Rop06,Rop07a}. This specific ignition pattern is
unlikely to happen at least in mildly turbulent convective core of the
white dwarf though \citep{Kuh06,Zin09,Zin11}.

This weakness of the pure deflagration model might be alleviated by a
later transition from deflagration to detonation (DDT) in the center
or at a certain radius \citep{Gam05}. The supersonic detonation wave
is expected to burn off the inhomogeneous unburned materials, release
enough energy to match classic SN Ia, and give stratified shells of
product elements. The synthetic light curves and spectra of the DDT
explosion agreed well with the observations of a normal SN Ia
\citep{Kas09}. It has been suspected that a DDT may take place when
the flame enters the distributed burning regime at the density of
$1\sim 3 \times 10^{7}\ \gpc$, where turbulence starts to deform
the internal flame structure and mix the hot ashes and fuel
\citep{Nie97}.  The existence of strong turbulence, with velocity
fluctuations larger than one fifth of local sound speed, is crucial
for a spontaneous detonation \citep{Woo09}, and \citet{Rop07}
suggested the possibility of these high turbulent velocities at the
corresponding density range from their deflagration simulations. We
plan to study the outcome of a later detonation following the current
deflagration calculations with 2 free parameters, the location and
time of the detonation wave initiation.

\acknowledgements

We thank Andy Aspden and Mike Zingale for helpful discussions
regarding the turbulent flame physics and the set up of problems in
hydrostatic equilibrium using the CASTRO code. The work at LBNL was
supported by the SciDAC Program of the DOE Office of High Energy
Physics and by the Applied Mathematics Program of the DOE Office of
Advance Scientific Computing Research under U.S. Department of Energy
under contract No. DE-AC02-05CH11231. The work at UCSC was supported
by the DOE SciDAC program, under grant No. DE-FC02-06ER41438 and by
the NASA Theory Program (NNX09AK36G).  Computer time for the
calculations in this paper was provided through a DOE INCITE award at
the Oak Ridge Leadership Computational Facility (OLCF) at Oak Ridge
National Laboratory, which is supported by the Office of Science of
the U.S. Department of Energy under Contract No. DE-AC05-00OR22725.
This research used resources of the National Energy Research
Scientific Computing Center, which is supported by the Office of
Science of the U.S. Department of Energy under Contract No.
DE-AC02-05CH11231. Early developmental work necessary to this study
was carried out on the Pleiades supercluster at UCSC. Pleiades was
purchased using a grant from the MRI Program of the NSF
(AST-0521566). Visualizations were performed using the VisIt
package. We thank Gunther Weber and Hank Childs for their assistance
with VisIt.

\newpage

\begin{table}
\caption{Final composition - deflagration ashes}
\begin{center}
\scalebox{0.9}{
\begin{tabular}{cccccccc}
\hline
\hline
{Log density} &
{He} &
{C}  &
{O}  &
{Ne} &
{Mg} &
{Si} &
{Fe} \\
{(g cm$^{-3}$})   &
   &
   &
   &
   &
   &
   &
    \\
\hline
    6.40  &    -   &  0.056 &  0.341 &  0.411 &  0.177 &  0.015 &    -   \\
    6.50  &    -   &  0.050 &  0.348 &  0.395 &  0.189 &  0.019 &    -   \\
    6.60  &    -   &  0.037 &  0.358 &  0.373 &  0.208 &  0.024 &    -   \\
    6.70  &    -   &  0.014 &  0.399 &  0.285 &  0.255 &  0.046 &    -   \\
    6.80  &    -   &  0.002 &  0.549 &  0.012 &  0.244 &  0.194 &    -   \\
    6.90  &    -   &    -   &  0.560 &    -   &  0.213 &  0.227 &    -   \\
    7.00  &    -   &    -   &  0.561 &    -   &  0.194 &  0.245 &    -   \\
    7.10  &    -   &    -   &  0.548 &    -   &  0.167 &  0.285 &    -   \\
    7.20  &    -   &    -   &  0.382 &    -   &  0.031 &  0.587 &    -   \\
    7.30  &    -   &    -   &    -   &    -   &    -   &  0.992 &  0.007 \\
    7.40  &  0.001 &    -   &    -   &    -   &    -   &  0.971 &  0.028 \\
    7.50  &  0.003 &    -   &    -   &    -   &    -   &  0.825 &  0.171 \\
    7.60  &  0.013 &    -   &    -   &    -   &    -   &  0.370 &  0.617 \\
    7.70  &  0.030 &    -   &    -   &    -   &    -   &  0.079 &  0.891 \\
    7.80  &  0.048 &    -   &    -   &    -   &    -   &  0.033 &  0.918 \\
    7.90  &  0.064 &    -   &    -   &    -   &    -   &  0.037 &  0.899 \\
    8.00  &  0.081 &    -   &    -   &    -   &    -   &  0.040 &  0.879 \\
    8.10  &  0.097 &    -   &    -   &    -   &    -   &  0.044 &  0.860 \\
    8.20  &  0.112 &    -   &    -   &    -   &    -   &  0.047 &  0.842 \\
    8.30  &  0.125 &    -   &    -   &    -   &    -   &  0.050 &  0.825 \\
    8.40  &  0.137 &    -   &    -   &    -   &    -   &  0.054 &  0.809 \\
    8.50  &  0.148 &    -   &    -   &    -   &    -   &  0.057 &  0.795 \\
    8.60  &  0.157 &    -   &    -   &    -   &    -   &  0.061 &  0.781 \\
    8.70  &  0.166 &    -   &    -   &    -   &    -   &  0.066 &  0.769 \\
    8.80  &  0.173 &    -   &    -   &    -   &    -   &  0.070 &  0.757 \\
    8.90  &  0.179 &    -   &    -   &    -   &    -   &  0.075 &  0.746 \\
    9.00  &  0.185 &    -   &    -   &    -   &    -   &  0.080 &  0.735 \\
    9.10  &  0.190 &    -   &    -   &    -   &    -   &  0.085 &  0.725 \\
    9.20  &  0.194 &    -   &    -   &    -   &    -   &  0.091 &  0.715 \\
    9.30  &  0.198 &    -   &    -   &    -   &  0.001 &  0.097 &  0.705 \\
    9.40  &  0.201 &    -   &    -   &    -   &  0.001 &  0.103 &  0.695 \\
    9.50  &  0.203 &    -   &    -   &    -   &  0.001 &  0.110 &  0.685 \\
    9.60  &  0.205 &    -   &    -   &    -   &  0.001 &  0.118 &  0.675 \\
    9.70  &  0.207 &    -   &  0.001 &    -   &  0.001 &  0.126 &  0.665 \\
    9.80  &  0.208 &    -   &  0.001 &    -   &  0.001 &  0.135 &  0.655 \\
    9.90  &  0.208 &  0.001 &  0.001 &    -   &  0.002 &  0.144 &  0.644 \\
    10.0  &  0.207 &  0.001 &  0.001 &    -   &  0.002 &  0.156 &  0.639 \\
\hline
\end{tabular}
\lTab{models1}
}
\end{center}
\end{table}

\clearpage

\begin{table}
\caption{Bulk properties - deflagration ashes}
\begin{center}
\scalebox{0.9}{
\begin{tabular}{cccccc}
\hline
\hline
{Log $\rho$} &
{$T_{9f}$} &
{$\rho_{7f}$}  &
{d$\rho_7$/d X$_{12}$}  &
{BE/A} &
{$\bar A$}  \\
{(g cm$^{-3}$)}   &
{ (10$^9$ K)}  &
{ (10$^7$ g cm$^{-3}$)}  &
{ (10$^7$ g cm$^{-3}$)}  &
{ (MeV/nucleon)}  &  
{ }   \\
\hline
 6.40  &  1.790 &   0.122 &   0.258 &    8.040 &   18.340 \\ 
 6.50  &  1.903 &   0.158 &   0.316 &    8.046 &   18.430 \\ 
 6.60  &  2.036 &   0.203 &   0.390 &    8.057 &   18.630 \\ 
 6.70  &  2.216 &   0.255 &   0.493 &    8.082 &   18.980 \\ 
 6.80  &  2.471 &   0.305 &   0.651 &    8.137 &   19.190 \\ 
 6.90  &  2.631 &   0.399 &   0.792 &    8.144 &   19.240 \\ 
 7.00  &  2.791 &   0.522 &   0.955 &    8.147 &   19.280 \\ 
 7.10  &  2.978 &   0.677 &   1.163 &    8.158 &   19.460 \\ 
 7.20  &  3.382 &   0.794 &   1.582 &    8.262 &   21.680 \\ 
 7.30  &  3.899 &   0.878 &   2.235 &    8.464 &   29.270 \\ 
 7.40  &  4.154 &   1.158 &   2.708 &    8.467 &   29.440 \\ 
 7.50  &  4.444 &   1.507 &   3.310 &    8.488 &   30.480 \\
 7.60  &  4.781 &   1.932 &   4.099 &    8.527 &   31.730 \\
 7.70  &  5.071 &   2.533 &   4.959 &    8.518 &   28.210 \\
 7.80  &  5.311 &   3.375 &   5.869 &    8.479 &   23.970 \\
 7.90  &  5.542 &   4.489 &   6.909 &    8.442 &   21.130 \\
 8.00  &  5.761 &   5.948 &   8.104 &    8.406 &   19.000 \\
 8.10  &  5.969 &   7.848 &   9.482 &    8.373 &   17.410 \\
 8.20  &  6.169 &  10.31 &  11.08 &    8.343 &   16.190 \\
 8.30  &  6.362 &  13.49 &  12.93 &    8.316 &   15.240 \\
 8.40  &  6.550 &  17.58 &  15.08 &    8.293 &   14.500 \\
 8.50  &  6.736 &  22.83 &  17.60 &    8.272 &   13.910 \\
 8.60  &  6.921 &  29.54 &  20.54 &    8.254 &   13.440 \\
 8.70  &  7.105 &  38.13 &  23.98 &    8.238 &   13.050 \\
 8.80  &  7.291 &  49.09 &  28.02 &    8.224 &   12.720 \\
 8.90  &  7.479 &  63.05 &  32.76 &    8.211 &   12.450 \\
 9.00  &  7.670 &  80.84 &  38.33 &    8.200 &   12.220 \\
 9.10  &  7.866 & 103.50 &  44.87 &    8.190 &   12.020 \\
 9.20  &  8.065 & 132.20 &  52.57 &    8.181 &   11.860 \\
 9.30  &  8.270 & 168.70 &  61.63 &    8.173 &   11.720 \\
 9.40  &  8.480 & 215.00 &  72.30 &    8.166 &   11.600 \\
 9.50  &  8.697 & 273.80 &  84.88 &    8.160 &   11.500 \\
 9.60  &  8.922 & 348.30 &  99.71 &    8.154 &   11.420 \\
 9.70  &  9.153 & 442.60 & 117.20 &    8.149 &   11.360 \\
 9.80  &  9.393 & 562.00 & 137.90 &    8.145 &   11.310 \\
 9.90  &  9.641 & 713.20 & 162.30 &    8.141 &   11.270 \\
 10.0  & 10.090 & 902.10 & 195.80 &    8.194 &   11.260 \\
\hline
\end{tabular}
\lTab{models2}
}
\end{center}
\end{table}

\clearpage
\begin{table}
\caption{Comparison with published flame models}
\begin{center}
\scalebox{0.9}{
\begin{tabular}{cccccc}
\hline
\hline
{Log density} &
{T$_{ash}$} &
{Pub. T$_{ash}$}  &
{$\Delta \rho/\rho$}  &
{Pub. $\Delta \rho/\rho$} &
{Ref.}  \\
{(g cm$^{-3}$)}   &
{(10$^9$ K)}  &
{(10$^9$K)}   &
{}   &
{}
\\
\hline
   6.78    &   2.42  &  2.45  &   0.514 & 0.527 &  Woo07 \\
   7.0     &   2.79  &  2.79  &   0.478 & 0.477 &  Woo07 \\
   7.398*  &   4.11  &  4.1   &   0.529 &  -    &  Woo11 \\
   7.477*  &   4.33  &  4.3   &   0.515 &  -    &  Woo11 \\
   8.0     &   5.76  &   -    &   0.405 & 0.426 &  Tim92 \\
   9.0     &   7.67  &   -    &   0.192 & 0.192 &  Tim92 \\
  10.0     &  10.1   &   -    &   0.098 & 0.094 &  Tim92 \\
\hline
\end{tabular}
\lTab{compare}
}
\end{center}
\end{table}

\clearpage

\begin{table}
\caption{Sample characteristics of nuclear statistical equilibrium}
\begin{center}
\scalebox{0.9}{
\begin{tabular}{ccccccccc}
\hline
\hline
{Log T} &
{Log density} &
{$Y_e$}   &
{X(He)}   &
{X(Si-Ca)}   &
{X(Fe group)}   &
{$\bar A$}&
{BE/A}    &
{d$Y_e$/dt}  \\
\hline
     9.80 &  9.30 & 0.500 & 1.601(-2) & 5.174(-2) & 0.9322 & 35.29 & 8.55635 & 2.617(-1) \\
     9.80 &  9.30 & 0.495 & 1.205(-2) & 4.097(-2) & 0.9469 & 40.79 & 8.60340 & 1.920(-1) \\
     9.80 &  9.30 & 0.490 & 9.798(-3) & 2.832(-2) & 0.9618 & 44.58 & 8.63952 & 1.313(-1) \\
     9.80 &  9.30 & 0.485 & 8.243(-3) & 1.645(-2) & 0.9752 & 47.26 & 8.66989 & 8.148(-2) \\
     9.80 &  9.30 & 0.480 & 7.000(-3) & 7.951(-3) & 0.9850 & 49.21 & 8.69495 & 4.668(-2) \\
     9.80 &  9.30 & 0.475 & 5.975(-3) & 3.575(-3) & 0.9904 & 50.57 & 8.71502 & 2.645(-2) \\
     9.80 &  9.30 & 0.470 & 5.042(-3) & 1.601(-3) & 0.9933 & 51.64 & 8.73201 & 1.449(-2) \\
     9.80 &  9.30 & 0.465 & 4.072(-3) & 7.147(-4) & 0.9952 & 52.58 & 8.74541 & 7.197(-3) \\
     9.80 &  9.30 & 0.460 & 3.159(-3) & 3.524(-4) & 0.9964 & 53.37 & 8.75220 & 3.470(-3) \\
     9.80 &  9.30 & 0.455 & 2.439(-3) & 1.989(-4) & 0.9973 & 53.96 & 8.75367 & 1.745(-3) \\
     9.80 &  9.30 & 0.450 & 1.841(-3) & 1.115(-4) & 0.9980 & 54.40 & 8.75201 & 8.264(-4) \\
     9.80 &  9.30 & 0.445 & 1.339(-3) & 5.506(-5) & 0.9986 & 54.74 & 8.74627 & 3.423(-4) \\
     9.80 &  9.30 & 0.440 & 1.001(-3) & 2.451(-5) & 0.9989 & 55.16 & 8.73595 & 1.219(-4) \\
     9.80 &  9.30 & 0.435 & 8.112(-4) & 1.029(-5) & 0.9991 & 55.70 & 8.72337 & 1.182(-5) \\
     9.80 &  9.30 & 0.430 & 7.294(-4) & 3.667(-6) & 0.9992 & 56.02 & 8.70984 & -5.946(-5) \\
     9.80 &  9.30 & 0.425 & 8.355(-4) & 8.179(-7) & 0.9991 & 55.10 & 8.69408 & -1.119(-4) \\
     9.80 &  9.30 & 0.420 & 1.933(-3) & 6.037(-8) & 0.9980 & 49.51 & 8.66726 & -1.310(-4) \\
\hline
\end{tabular}
\lTab{nse}
}
\end{center}
\end{table}

\clearpage

\begin{table}
\caption{Flame adjustment parameters, $f_w$,$f_v$, at different densities}
     \label{tab:laminar_parameter}
\begin{center}
\scalebox{0.9}{
\begin{tabular}{c|cc}
{log($\rho$)} &
{$f_w$}       &
{$f_v$}       
\\
\hline
6.90 & 2.40 & 2.90  \\
7.00 & 2.00 & 2.20  \\
7.15 & 1.80 & 1.95  \\
7.30 & 1.40 & 1.55  \\
7.50 & 1.00 & 1.20  \\
7.70 & 1.00 & 1.05  \\ 
7.85 & 1.20 & 1.10  \\
8.00 & 1.20 & 1.15  \\
8.30 & 1.60 & 1.30  \\
8.50 & 1.80 & 1.40  \\
8.70 & 2.00 & 1.60  \\
8.85 & 2.20 & 1.70  \\
9.00 & 2.40 & 1.75  \\
9.20 & 2.80 & 2.00  \\ 
9.30 & 3.00 & 2.20  \\
9.40 & 3.40 & 2.30  \\ 
9.45 & 3.40 & 2.30  \\
\end{tabular}
}
\end{center}
\end{table}

\clearpage

\begin{table}
\caption{Mass of chemical products and the total released energy}
\begin{center}
\scalebox{0.7}{
\begin{tabular}{ccccccccccccc}
\hline
\hline
{Model} &
{M(C)}        &
{M(O)}        &
{M(He)}       &
{M(Ne)}       &
{M(Mg)}       &
{M(IME)}      &
{M("Fe")}     &
{M($\Feff$)}  &
{M($\Fefs$)}  &
{M($\Nifs$)}  &
{M($\Nife$)}  &
{E$_{nuc}$} 
\\
    &
{($\Msun$)}   &
{($\Msun$)}   &
{($\Msun$)}   &
{($\Msun$)}   &
{($\Msun$)}   &
{($\Msun$)}   &
{($\Msun$)}   &
{($\Msun$)}   &
{($\Msun$)}   &
{($\Msun$)}   &
{($\Msun$)}   &
{(10$^{51}$ ergs)}  
\\
\hline
   A50 & 0.17  & 0.24  &  0.04  &  0.006 &  0.03  &  0.17  &  0.73  &  0.12  & 0.0012 &  0.56  &  0.04  &  1.26 \\
  A100 & 0.22  & 0.30  &  0.05  &  0.005 &  0.03  &  0.16  &  0.63  &  0.11  & 0.0005 &  0.48  &  0.04  &  1.15 \\
  A200 & 0.21  & 0.29  &  0.03  &  0.006 &  0.03  &  0.17  &  0.65  &  0.11  & 0.0015 &  0.50  &  0.04  &  1.16 \\
  B100\tablenotemark{a} & 0.48 & 0.48 & 0.04& $4.0(-7)$& 
  $5.5(-5)$& 0.02 & 0.37 & - & - & - & - & - \\
  BT\tablenotemark{a} & 0.47 & 0.47 & 0.04 & $1.5(-6)$& 
  $6.2(-5)$& 0.02 & 0.38 & - & - & - & - & - \\
  b150\_3d\tablenotemark{b} & 0.28 & 0.28 &  -   &  -      &  -      &  0.17  &  0.67  &   -     & -
      &  -     &   -     &  1.18 \\
  b250\_3d\tablenotemark{b} & 0.31 & 0.31 &  -   &   -     &  -      &  0.16  &  0.61  &  -      &   -      &  -     &  -      &  1.09 \\
 SNOB\tablenotemark{c} & 0.18  & 0.18   &   -    &  -      &   -     &  0.43  &  0.61 &     -    &   -      & 0.33  &     -   &  1.31 \\
\hline
\tablenotetext{a}{Not run to burning completion --- only out to 0.74 s; 
  should only be used for comparison between B series simulations}
\tablenotetext{b}{\cite{Rop06}}
\tablenotetext{c}{\cite{Rop07a}}
\end{tabular}
\lTab{products}
}
\end{center}
\end{table}



\clearpage

\begin{figure}
\centering
\includegraphics[angle=90,width=6.0in]{\figurepath 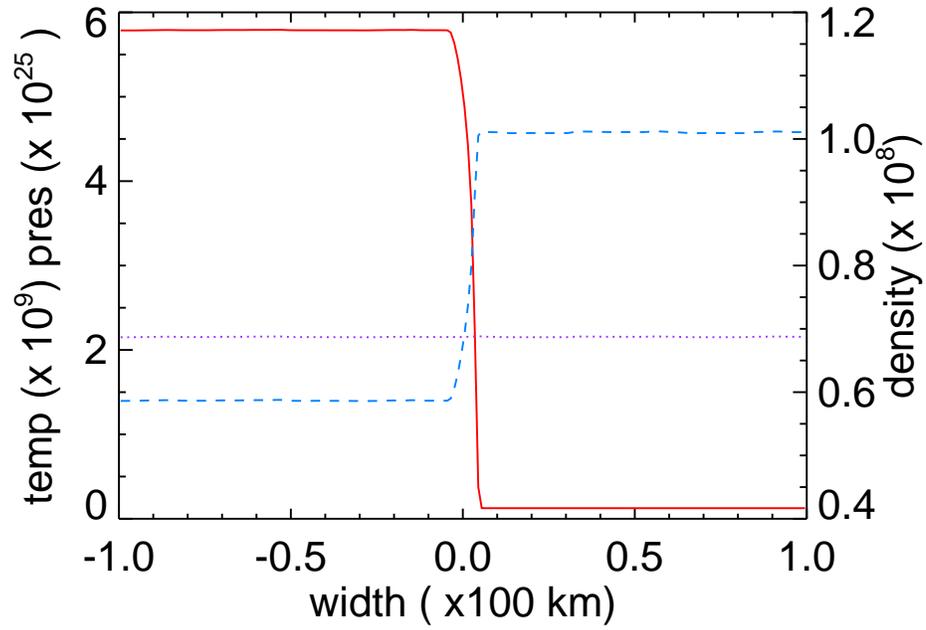}
\caption[Density, temperature and pressure profiles across the flame]
        {Density (dashed), temperature (solid) and pressure (dotted)
          profiles across the 1D laminar flame surface.  \lFig{f1}}
\end{figure}

\clearpage

\begin{figure}
\centering
\includegraphics[angle=90,width=6.0in]{\figurepath 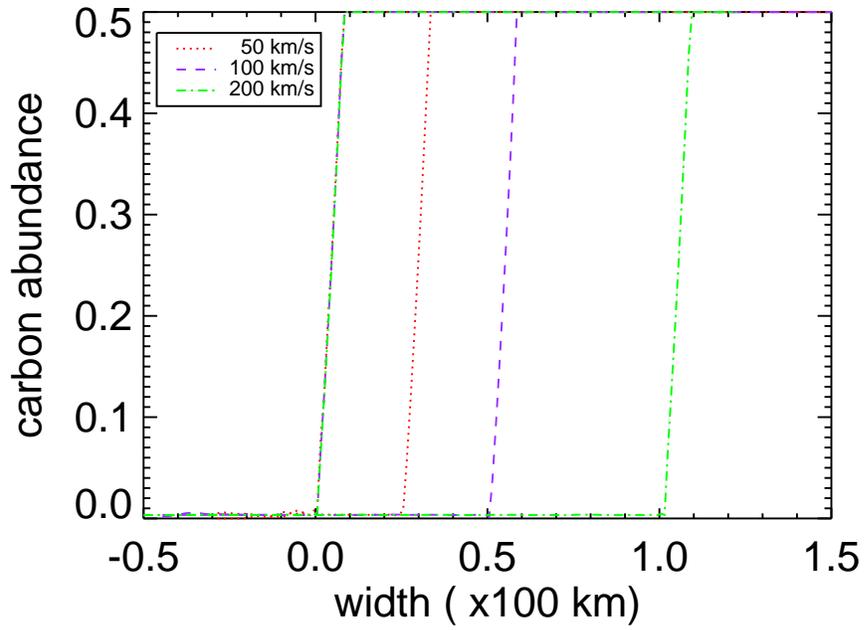}
\caption[1D Laminar flame propagation for flame speeds of 50, 100 and 200]
        {1D laminar flame propagation for flame speeds of 50, 100 and 200 \kms. 
        Flame fronts at 0.3 s are put at the same location, and 
        flame fronts for different prescribed speeds at 0.8 s are shown.
  \lFig{f2}}
\end{figure}

\clearpage

\begin{figure*}
\centering
\includegraphics[width=0.495\textwidth,clip=true]{\figurepath 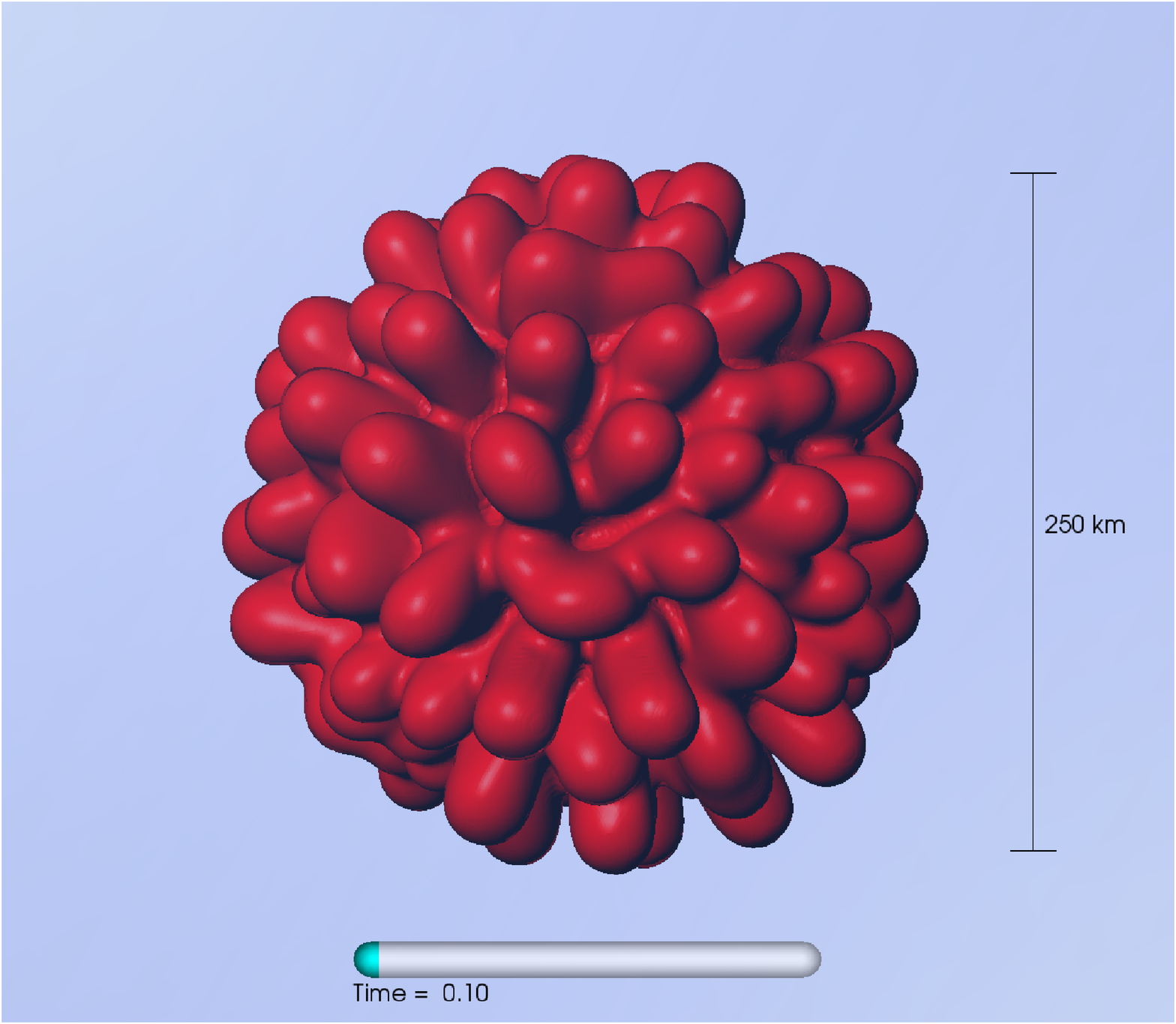}
\hfill
\includegraphics[width=0.495\textwidth,clip=true]{\figurepath 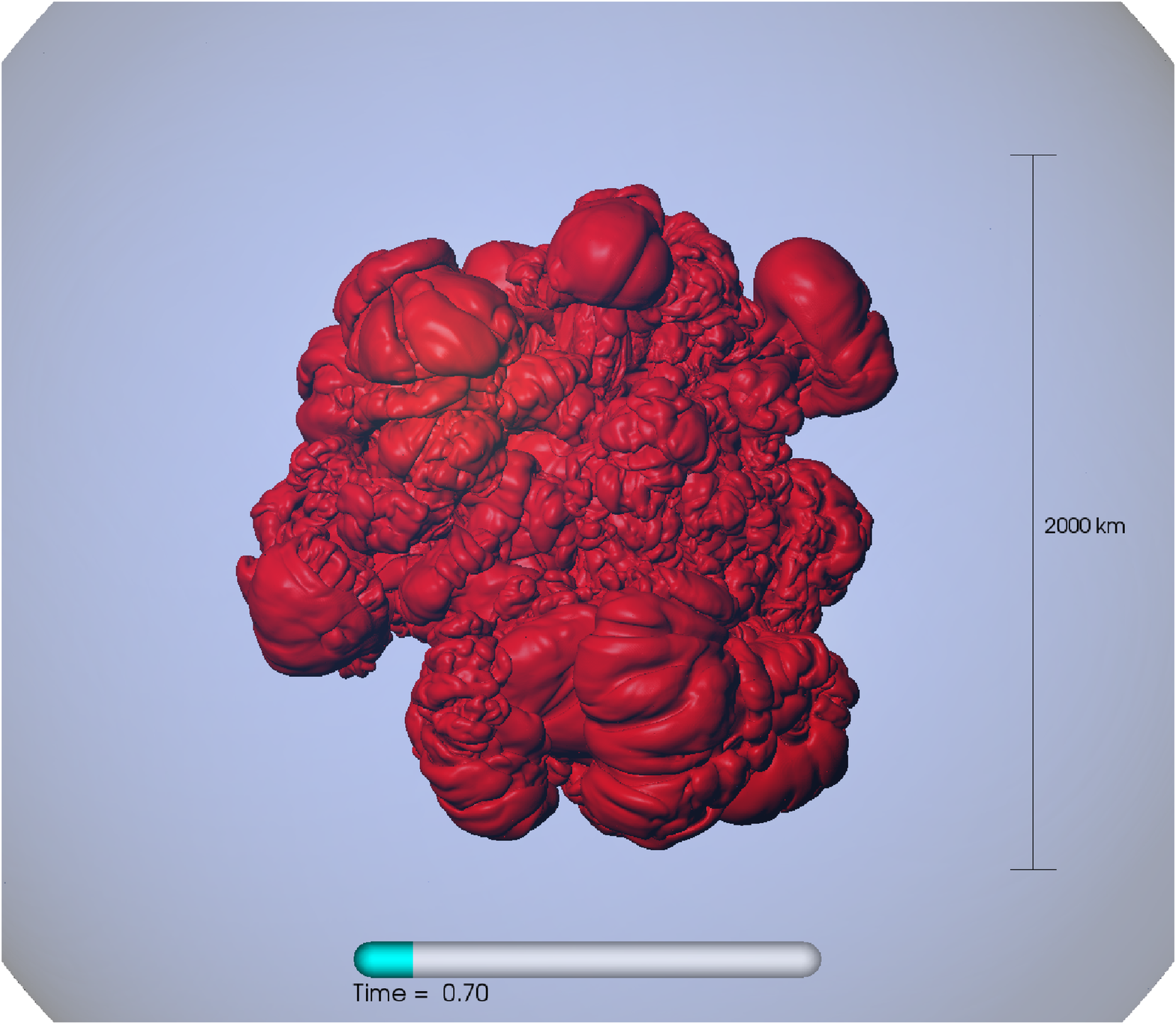}
\vskip 24pt
\includegraphics[width=0.495\textwidth,clip=true]{\figurepath 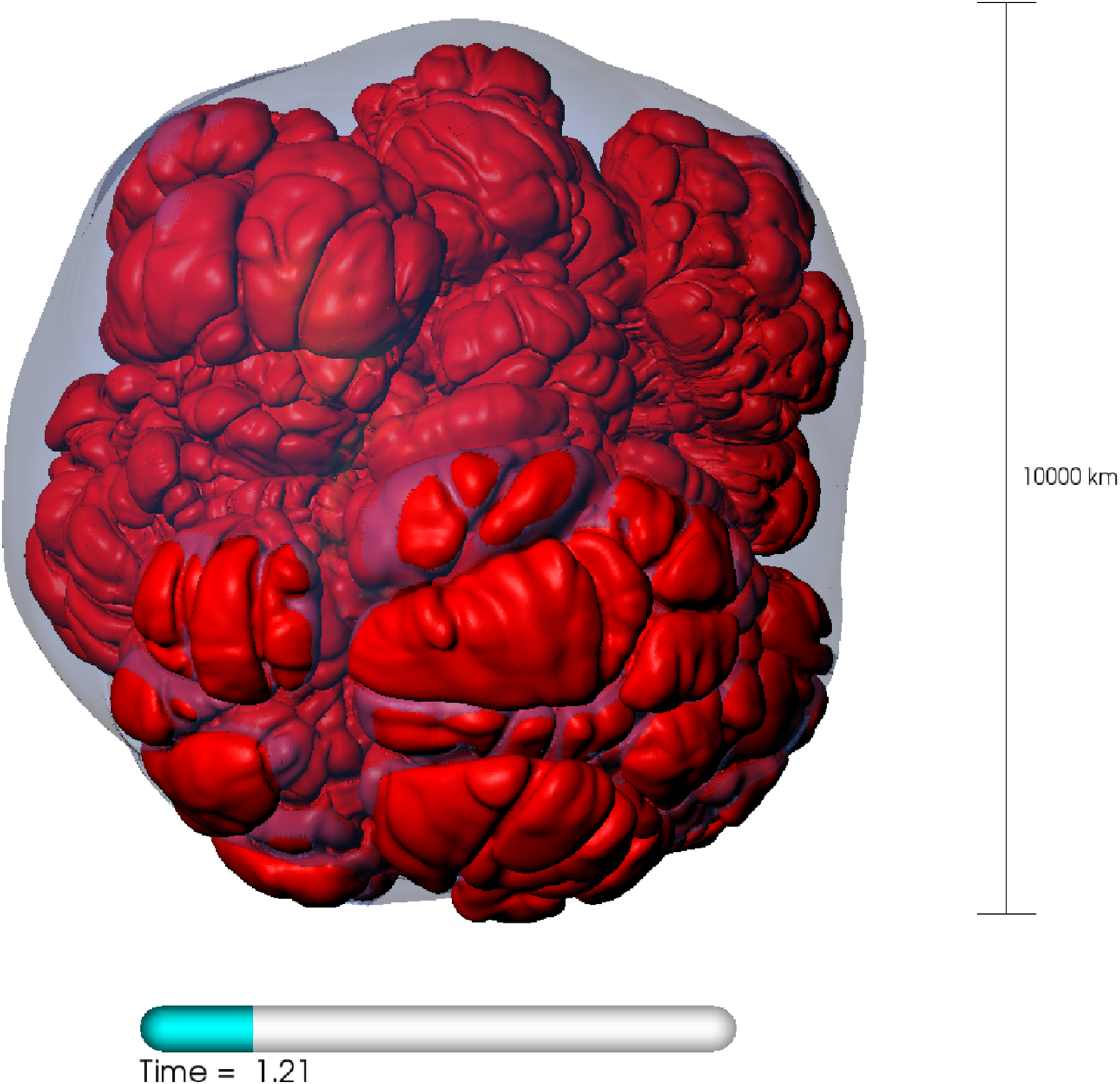}
\hfill
\includegraphics[width=0.495\textwidth,clip=true]{\figurepath 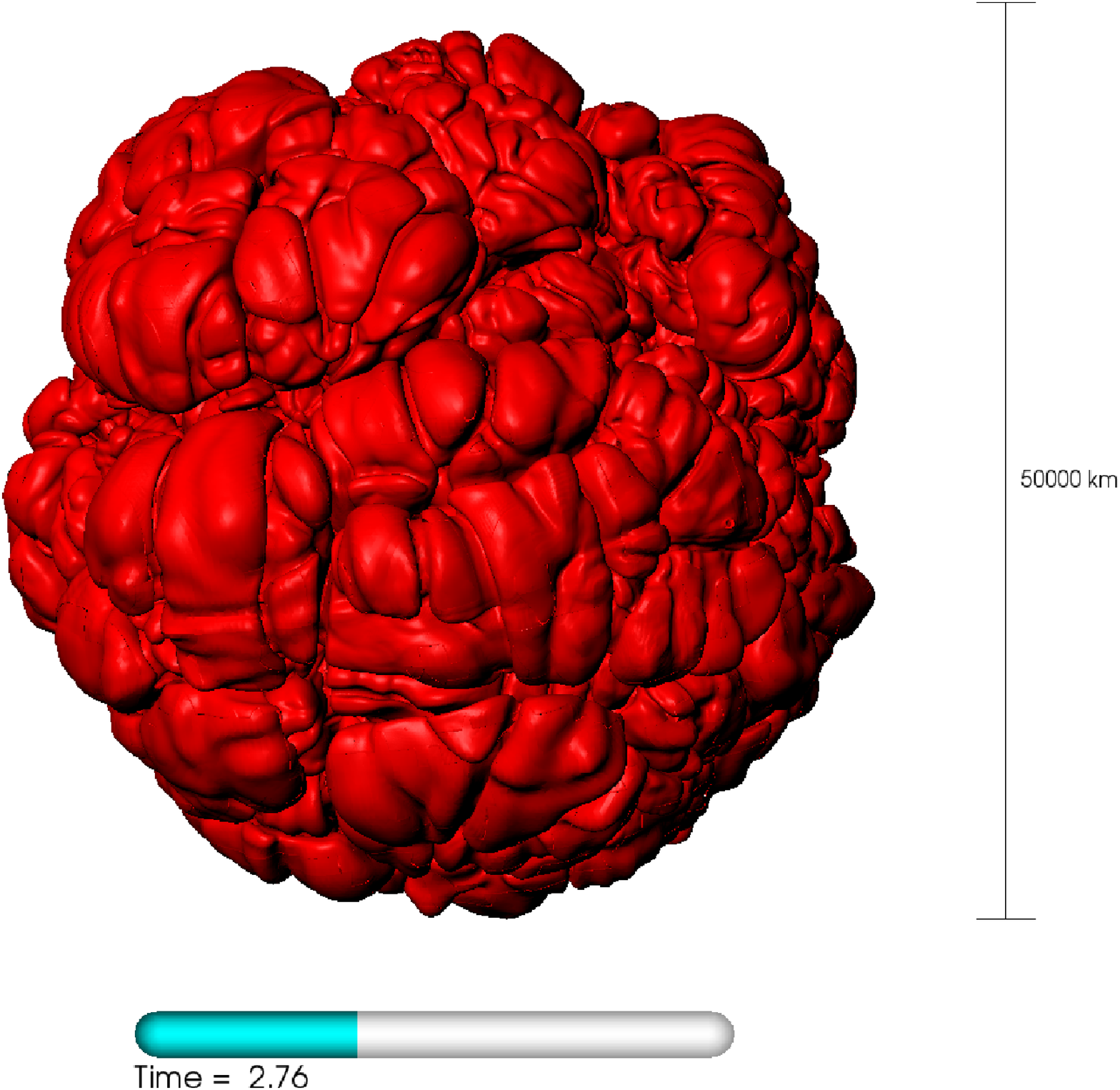}
\vskip 24pt
\caption[Contour plots of the ash-fuel interfaces for flame speed 100
] {Contour plots of the ash-fuel interfaces showing the development of
  the thermonuclear deflagration in the carbon-oxygen white dwarf in
  the simulation with flame speed parameter 100 km/s, model A100. Star
  time and length scale are shown on each plot. Red color indicates
  the interface, and light blue color indicates the star surface where
  density equals $10^{6}$ $\mathrm{g~cm}^{-3}$.  \lFig{f3}}
\end{figure*}

\clearpage

\begin{figure}
\centering
\includegraphics[angle=90,width=6.0in]{\figurepath 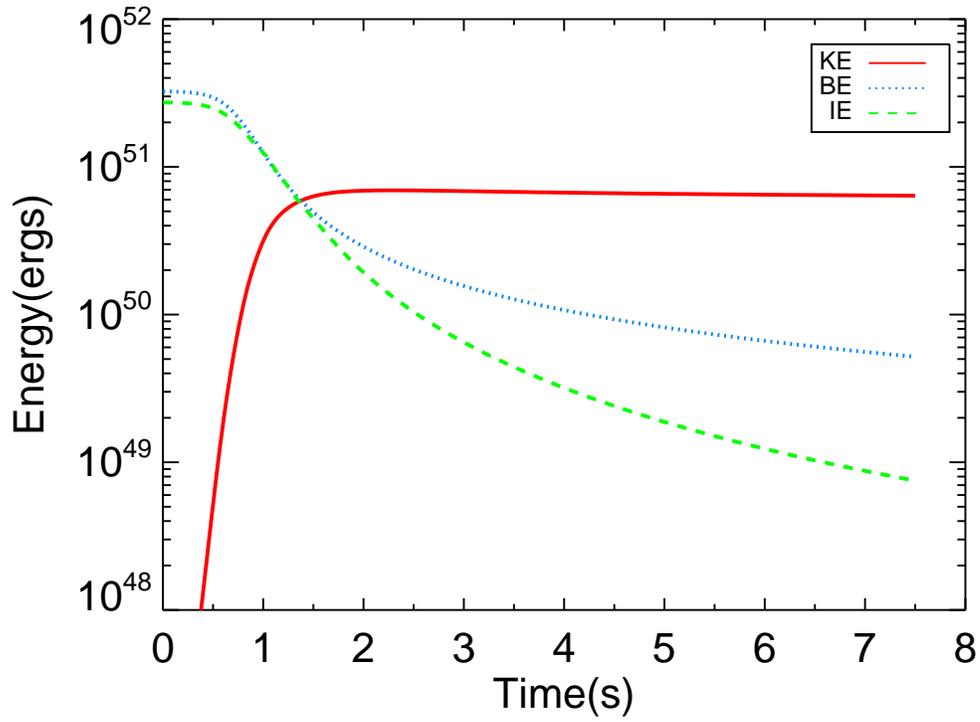}
\caption[kinetic energy, binding energy and internal energy evolution]
        {Kinetic energy, binding energy and internal energy evolution 
         for model A100.
  \lFig{f4}}
\end{figure}

\clearpage
\begin{figure*}
\centering
\includegraphics[angle=90,width=0.495\textwidth,clip=true]{\figurepath 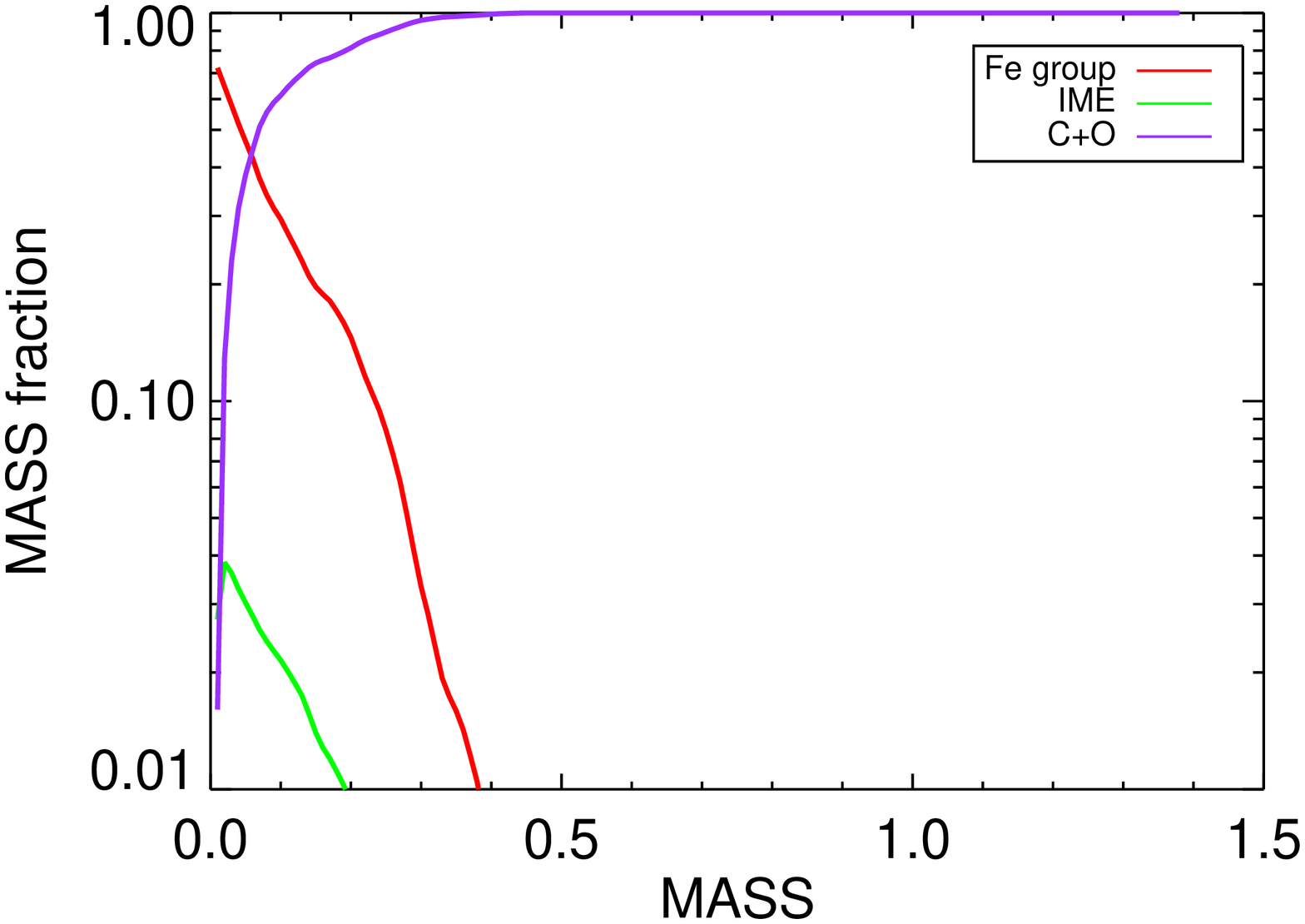}
\hfill
\includegraphics[angle=90,width=0.495\textwidth,clip=true]{\figurepath 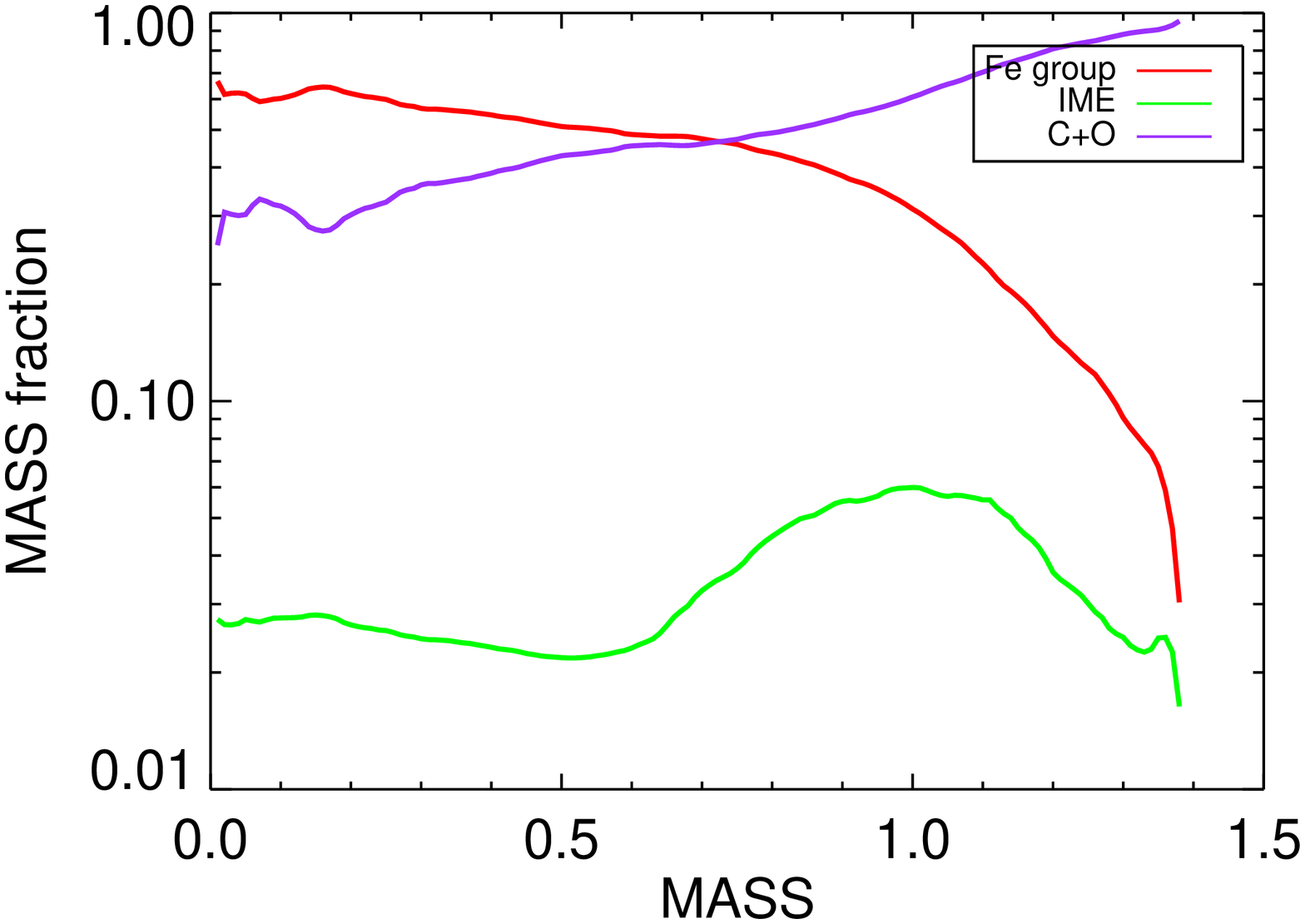}
\vskip 24pt
\includegraphics[angle=90,width=0.495\textwidth,clip=true]{\figurepath 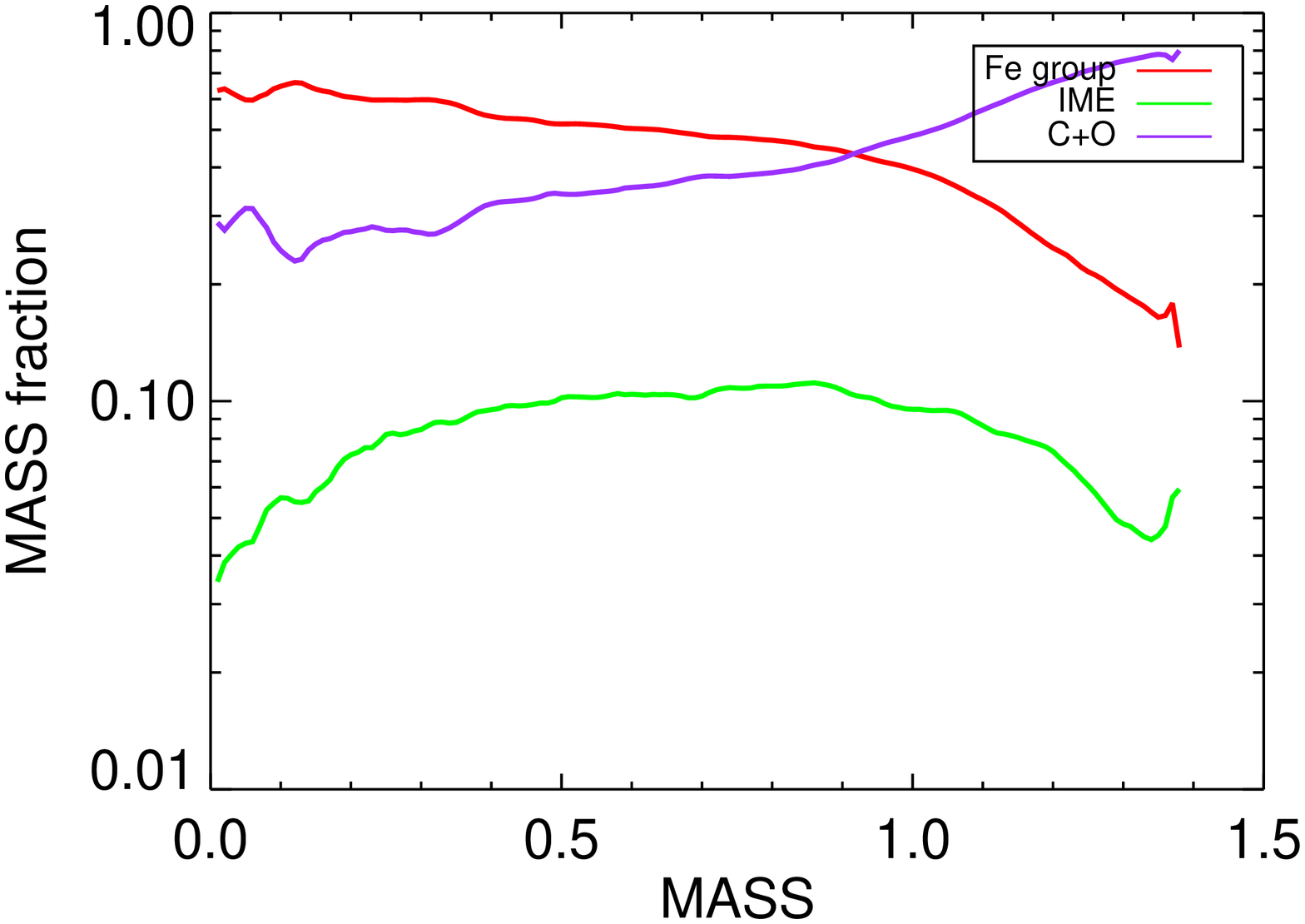}
\hfill
\includegraphics[angle=90,width=0.495\textwidth,clip=true]{\figurepath 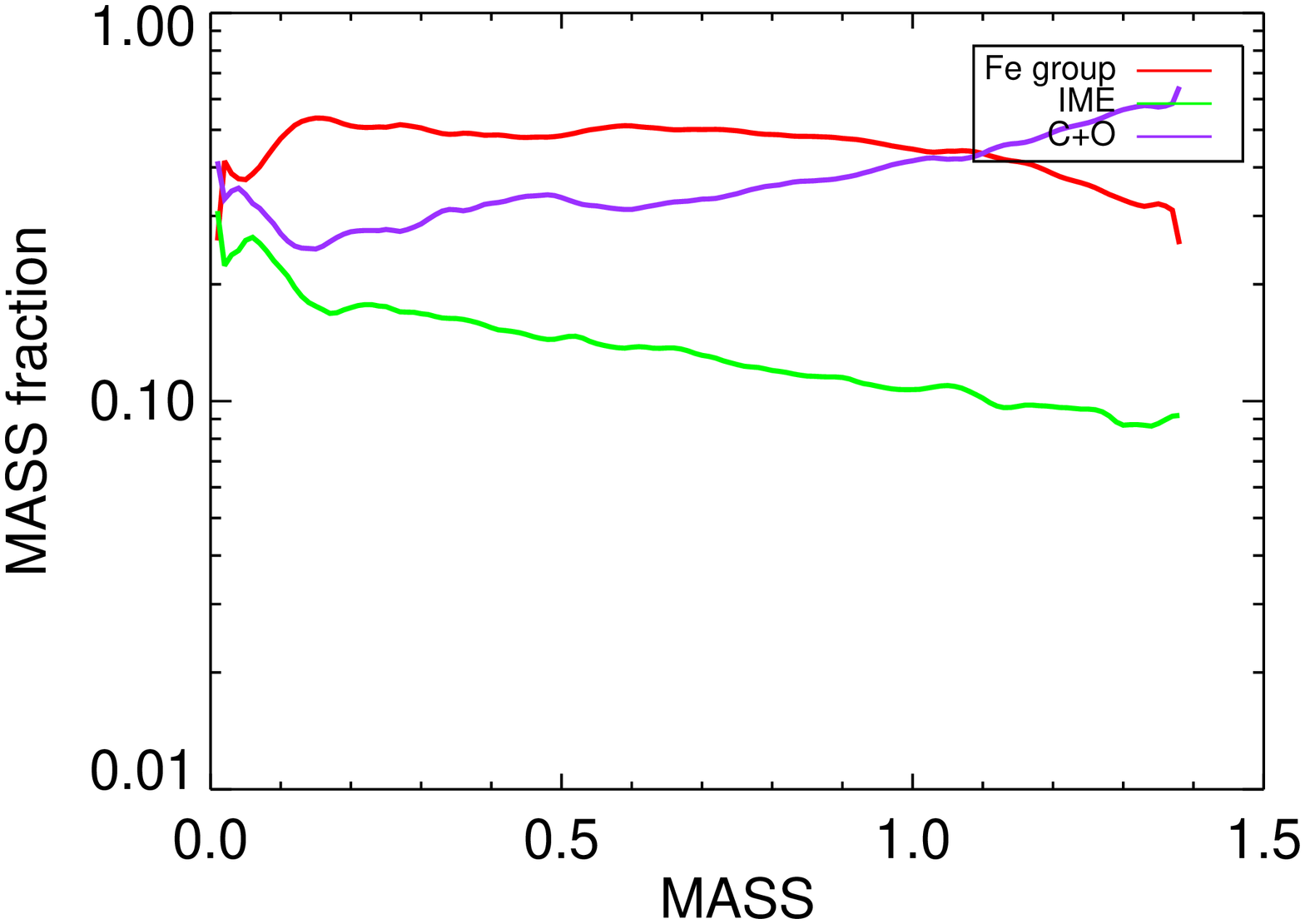}
\caption[Mass fraction of iron, IME and unburned material at 0.5 s, 1 s, 1.2 s and 1.7 s]
        {Mass fraction of iron, IME and unburned material at 0.5 s, 1 s, 1.2 s and 1.7 s 
         for model A100.
  \lFig{f5}}
\end{figure*}

\clearpage
\begin{figure*}
\centering
\includegraphics[width=0.32\textwidth,clip=true]{\figurepath 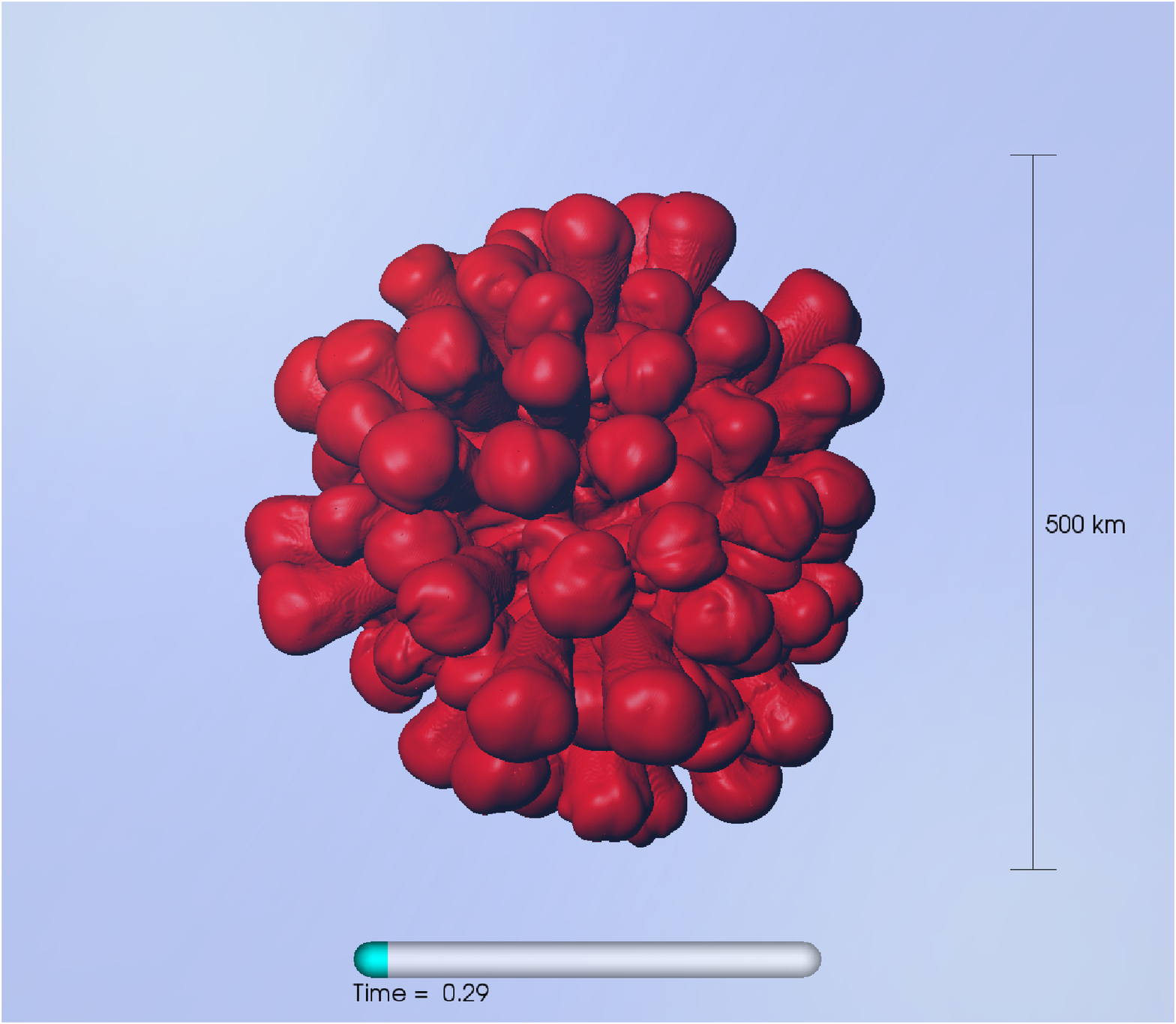}
\includegraphics[width=0.32\textwidth,clip=true]{\figurepath 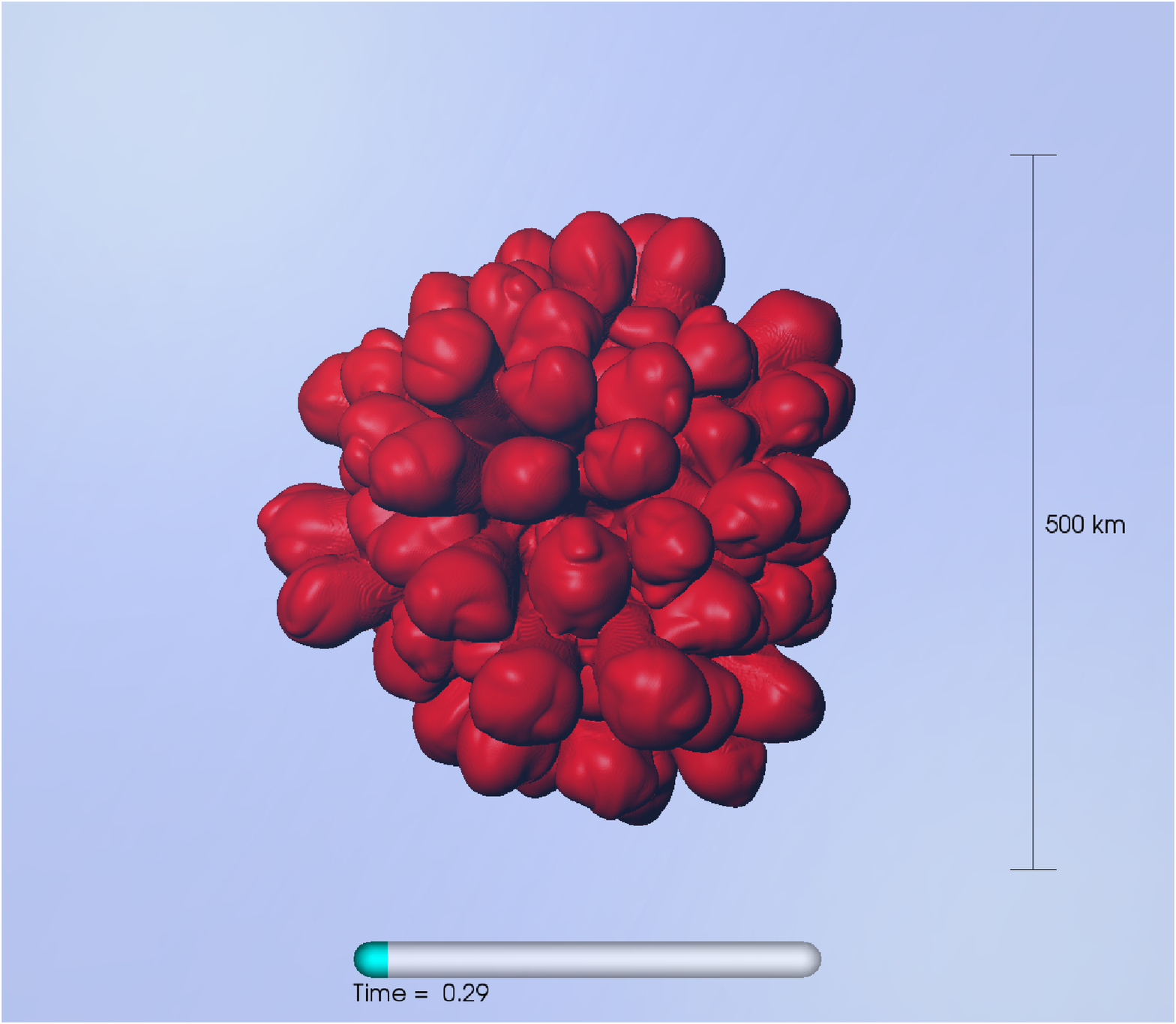}
\includegraphics[width=0.32\textwidth,clip=true]{\figurepath 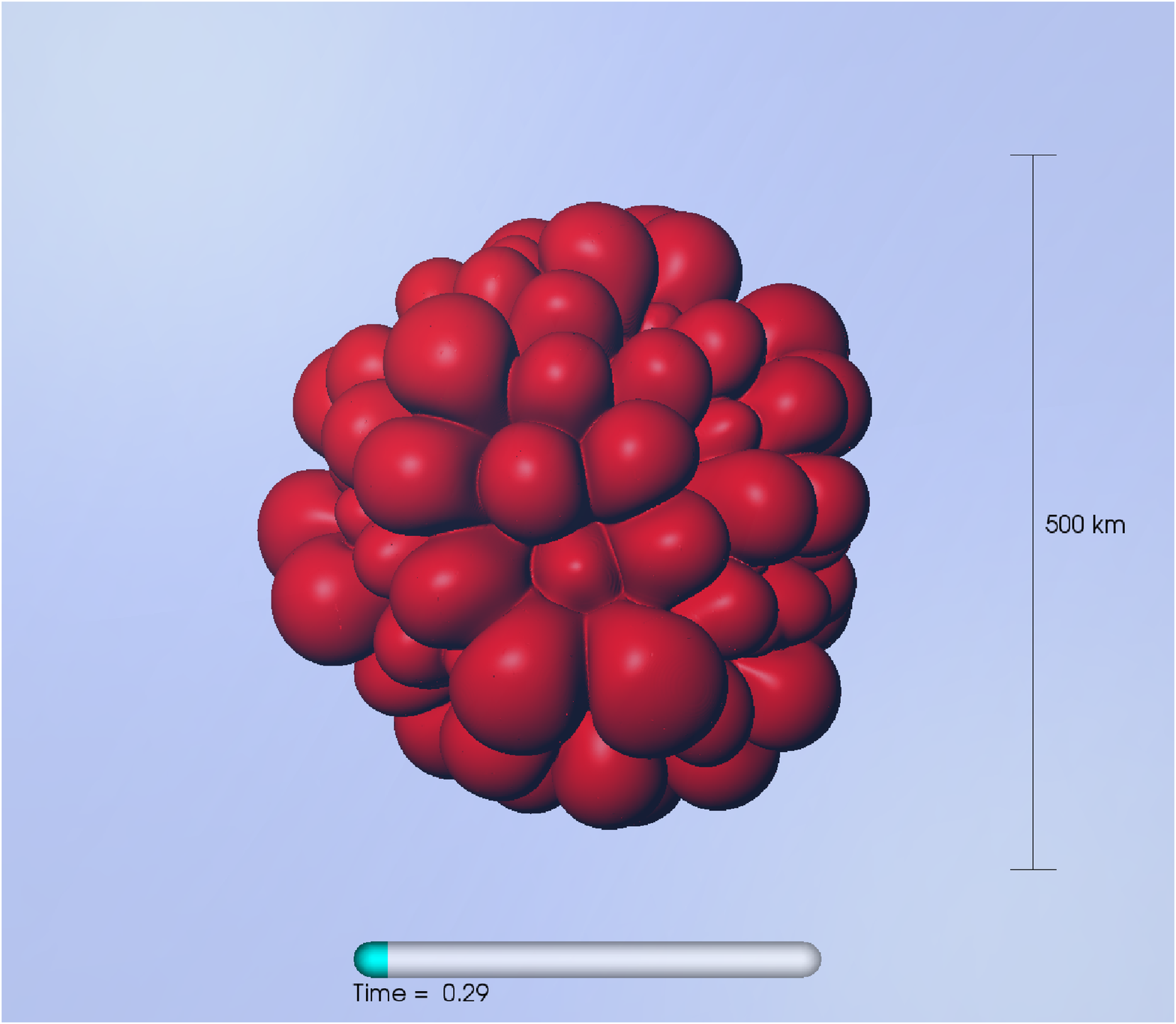}
\hfill
\vskip 24pt
\includegraphics[width=0.32\textwidth,clip=true]{\figurepath 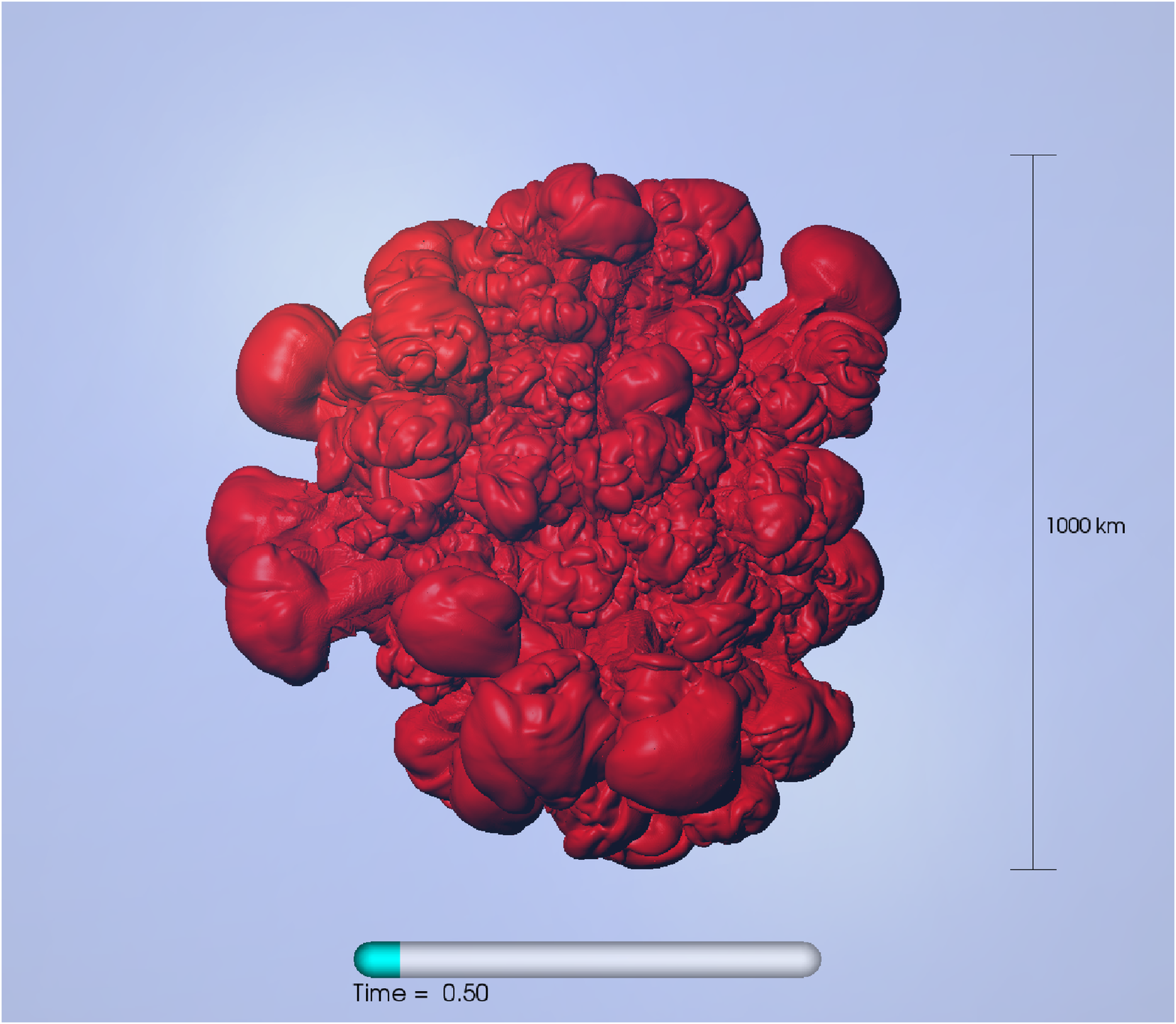}
\includegraphics[width=0.32\textwidth,clip=true]{\figurepath 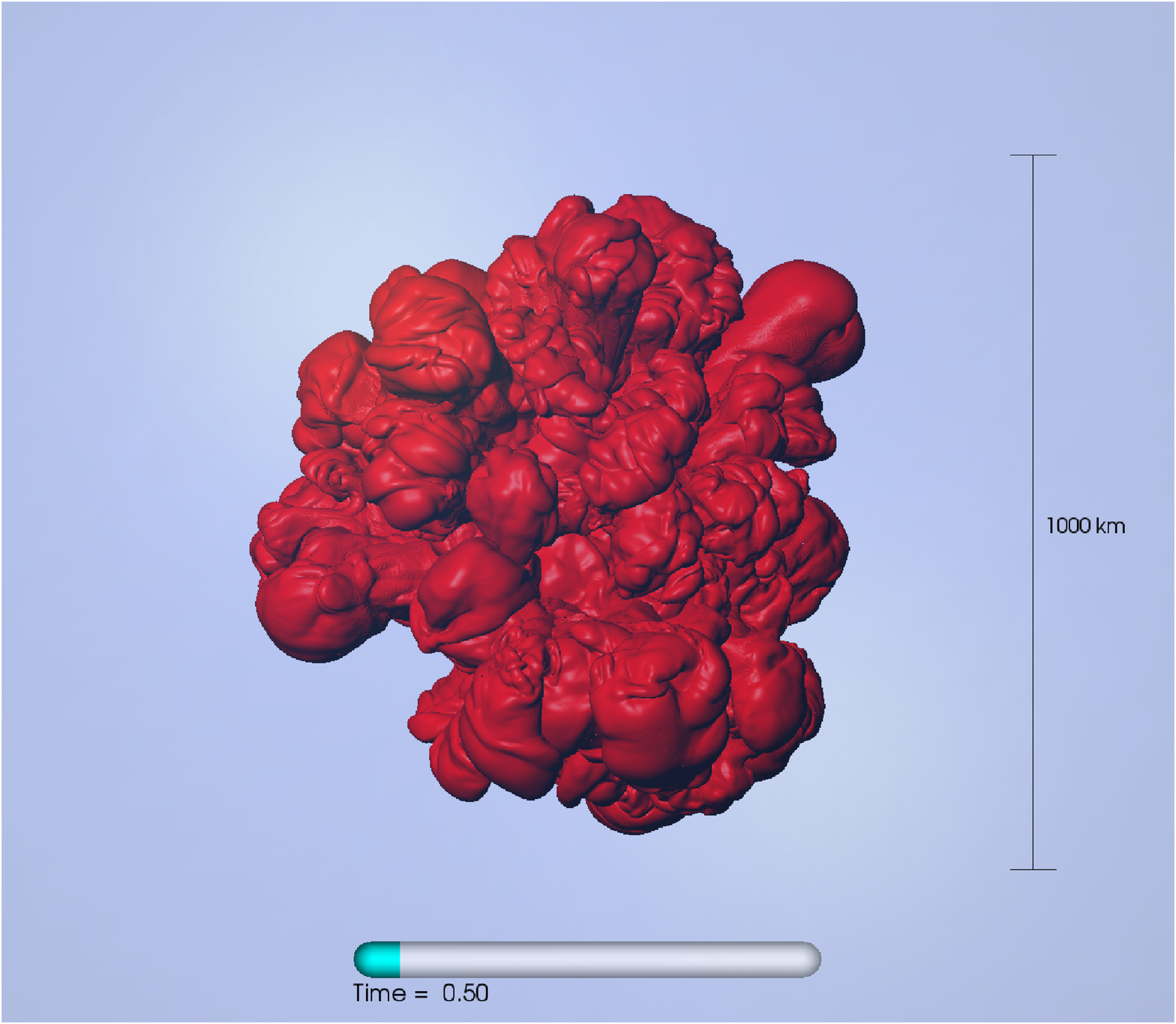}
\includegraphics[width=0.32\textwidth,clip=true]{\figurepath 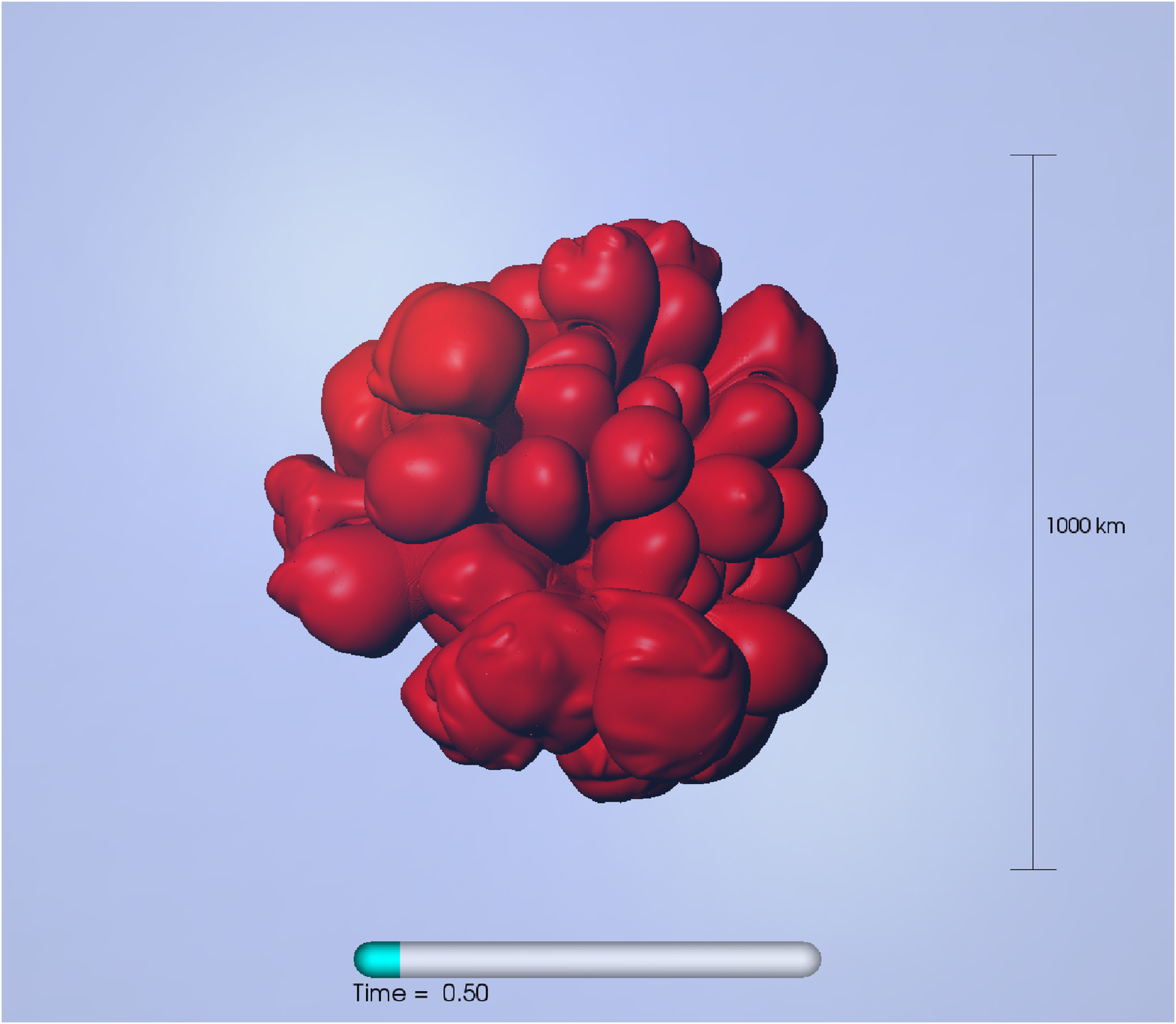}
\hfill
\vskip 24pt
\includegraphics[width=0.32\textwidth,clip=true]{\figurepath 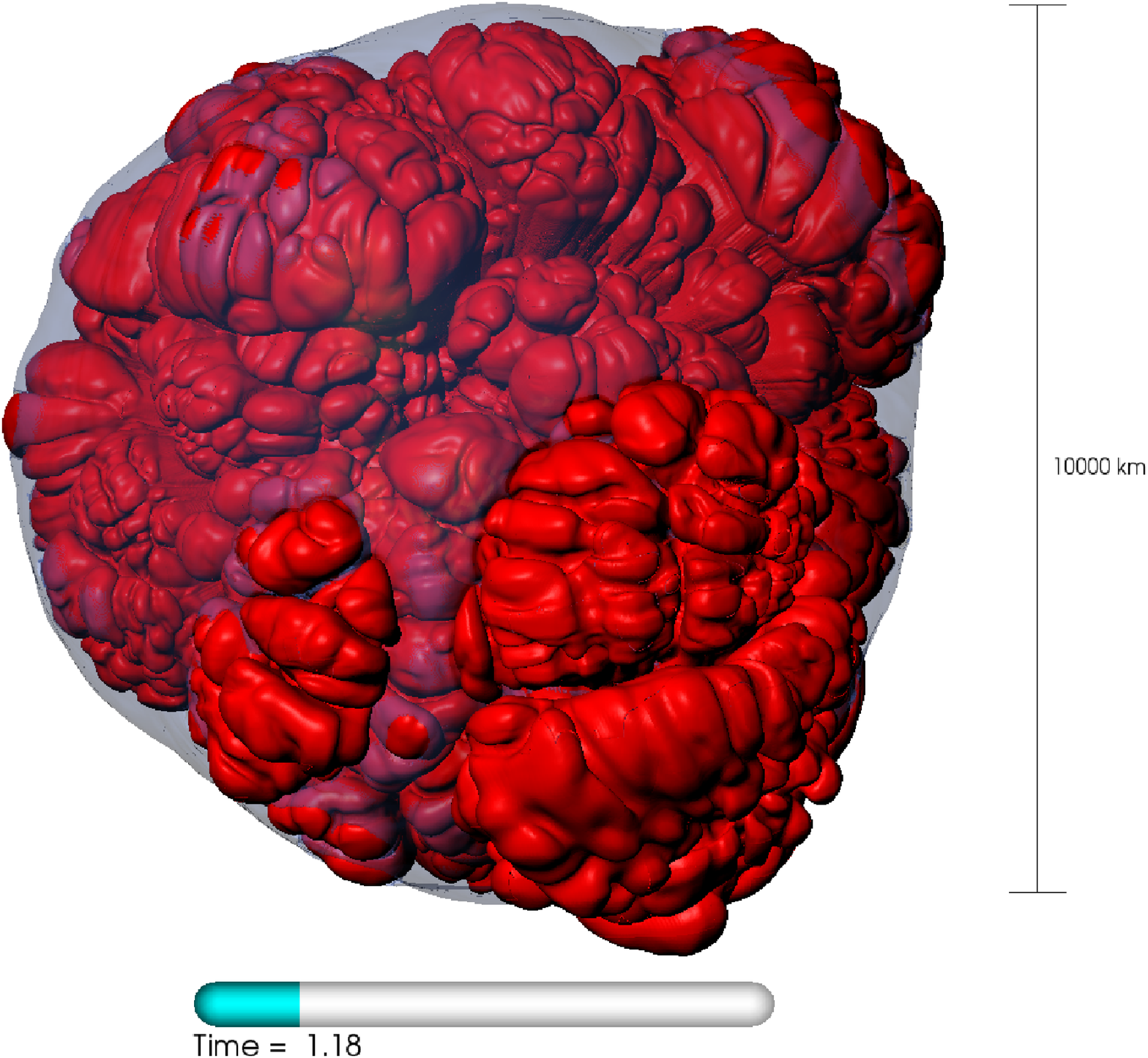}
\includegraphics[width=0.32\textwidth,clip=true]{\figurepath 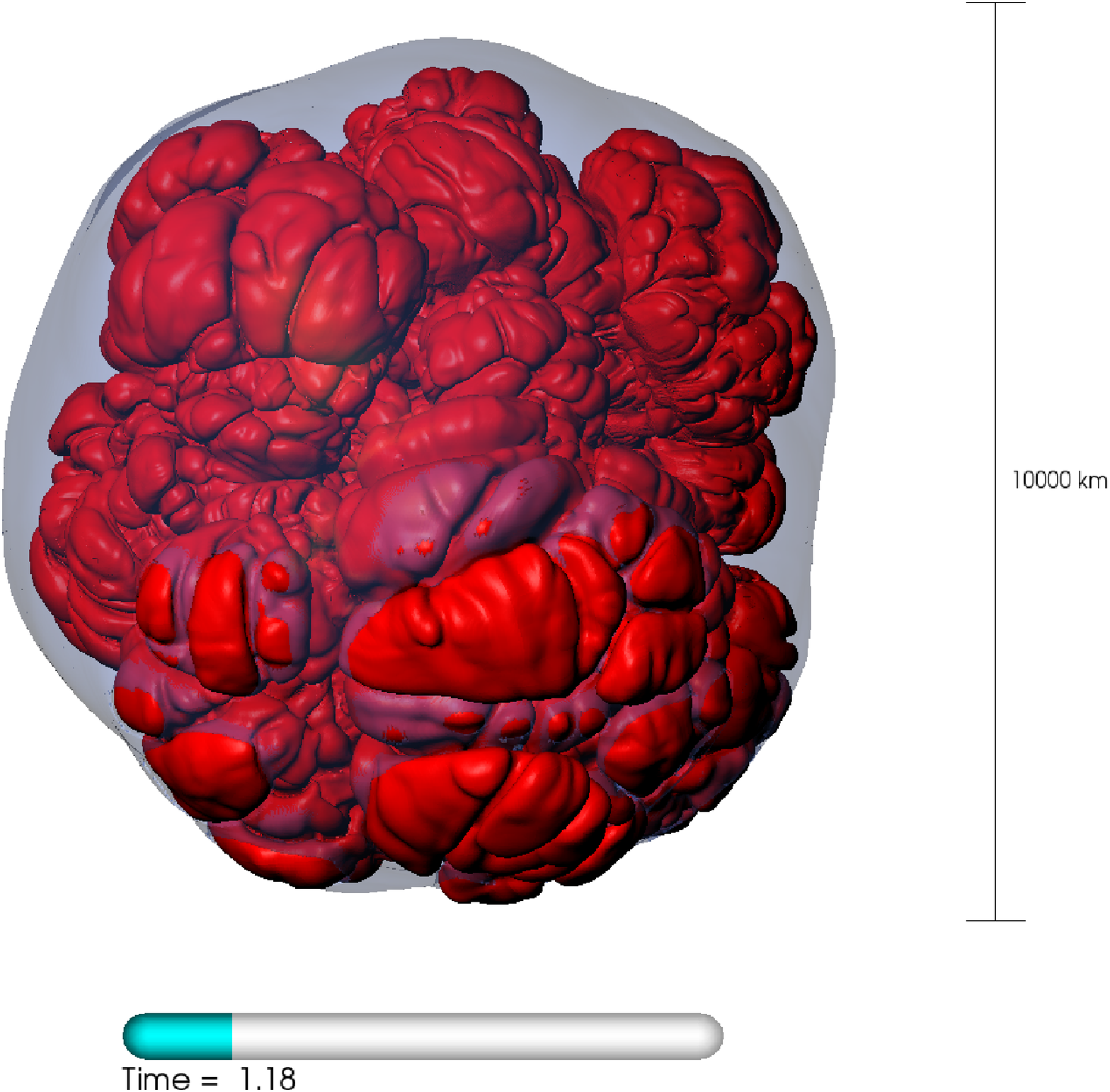}
\includegraphics[width=0.32\textwidth,clip=true]{\figurepath 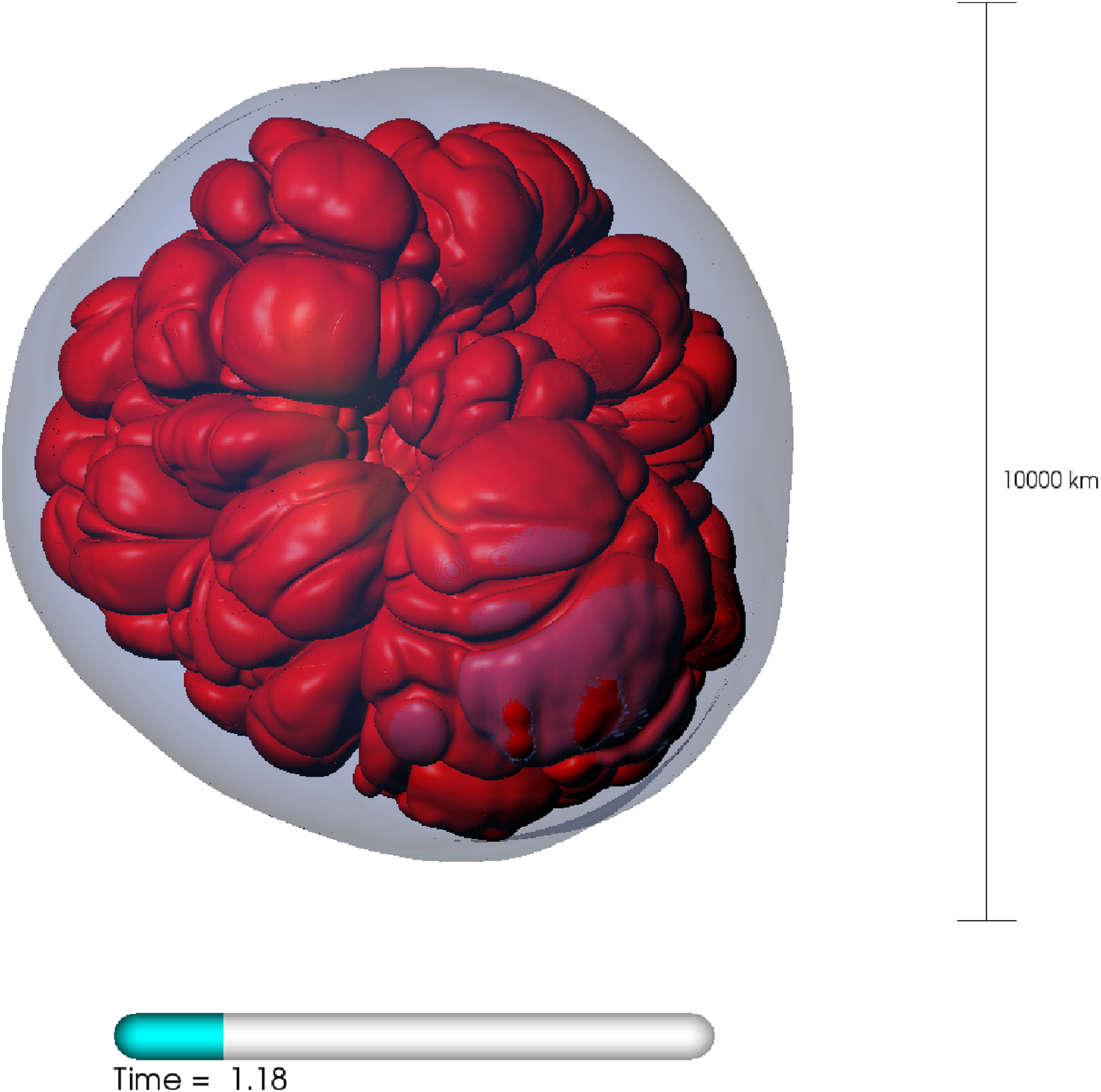}
\hfill
\vskip 24pt
\caption[Contour plots of the burning interfaces for 3 flame speeds models]
        {Contour plots of the burning interfaces for 3 different prescribed 
         flame speed models, A50, A100 and A200 (from left to right) at the star 
         time of 0.3 s, 0.5 s and 1.2 s (from top to bottom).
  \lFig{f6}}
\end{figure*}

\clearpage
\begin{figure}
\centering
\includegraphics[angle=90,width=6.0in]{\figurepath 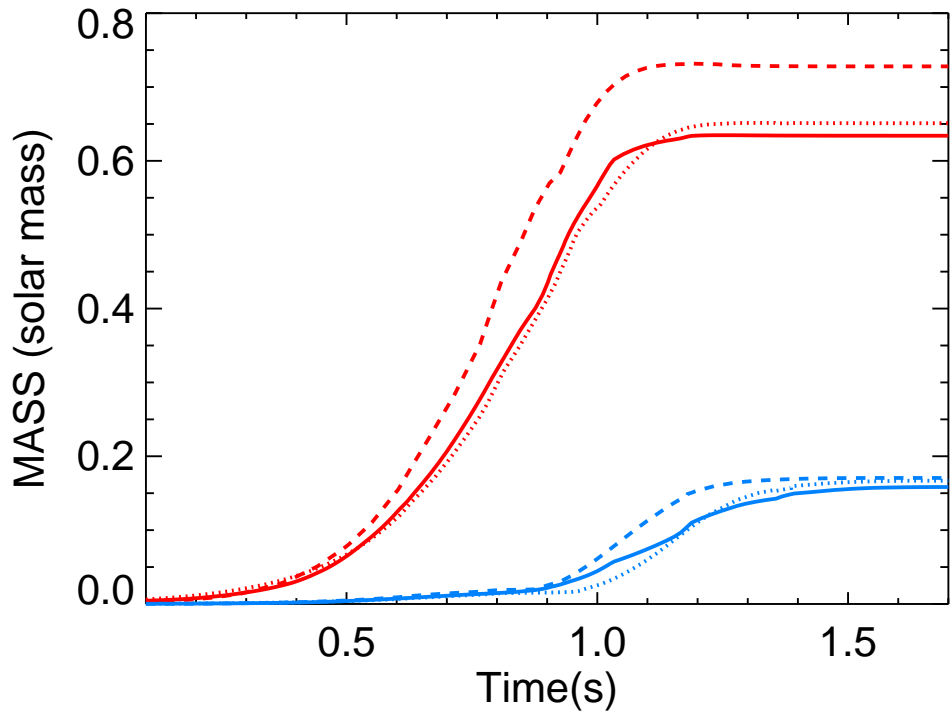}
\caption[Mass of produced iron group elements (red) and IME (blue) for 3 flame
  speed models] {Mass of produced iron group elements (red) and IME (blue) with
  time for 3 different prescribed flame speed models. The solid lines
  correspond to model A100, the dashed lines to model A50, and the
  dotted lines to A200.  \lFig{f7}}
\end{figure}

\clearpage
\begin{figure*}
\centering
\includegraphics[width=0.4\textwidth,clip=true]{\figurepath 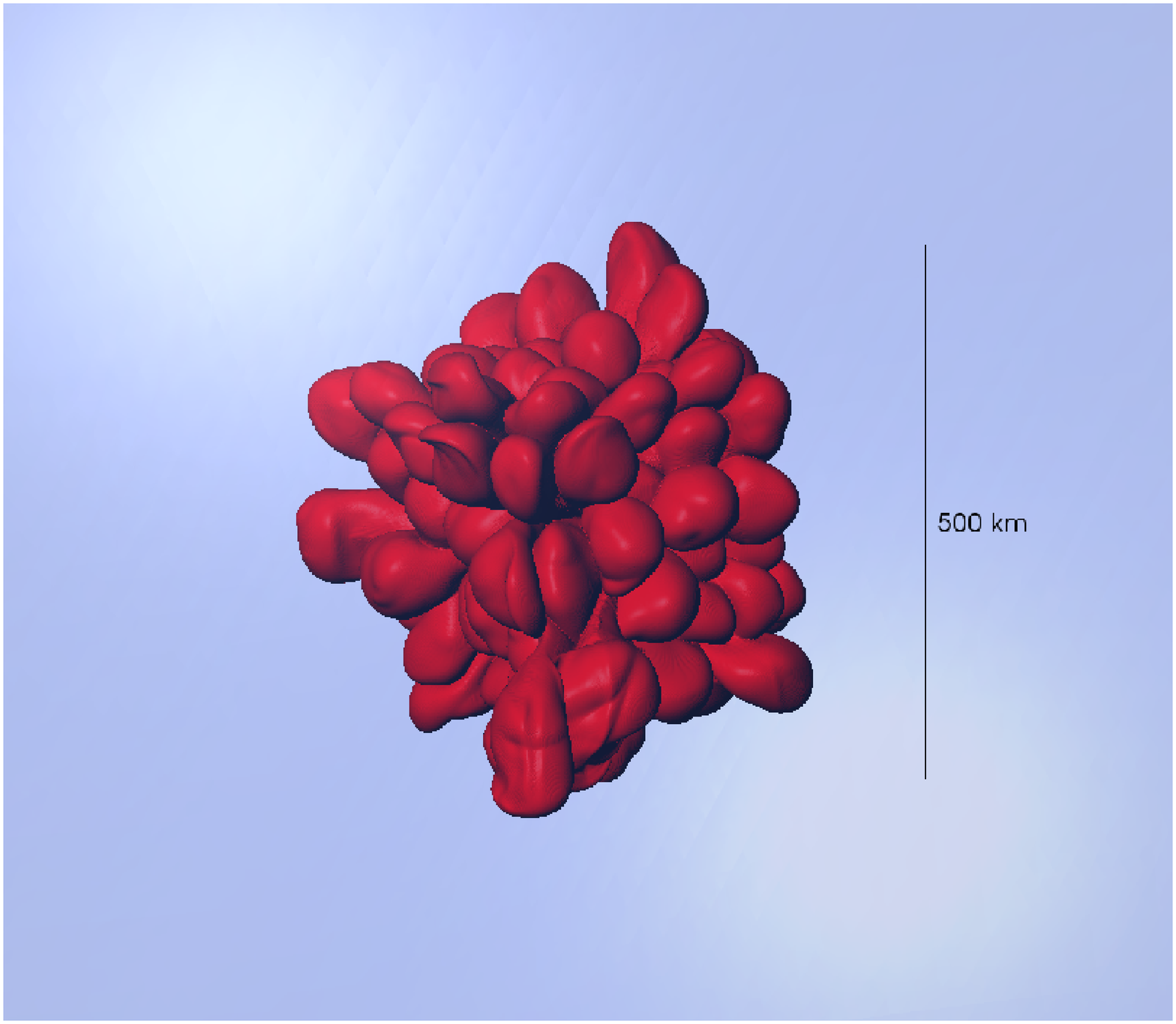}
\includegraphics[width=0.4\textwidth,clip=true]{\figurepath 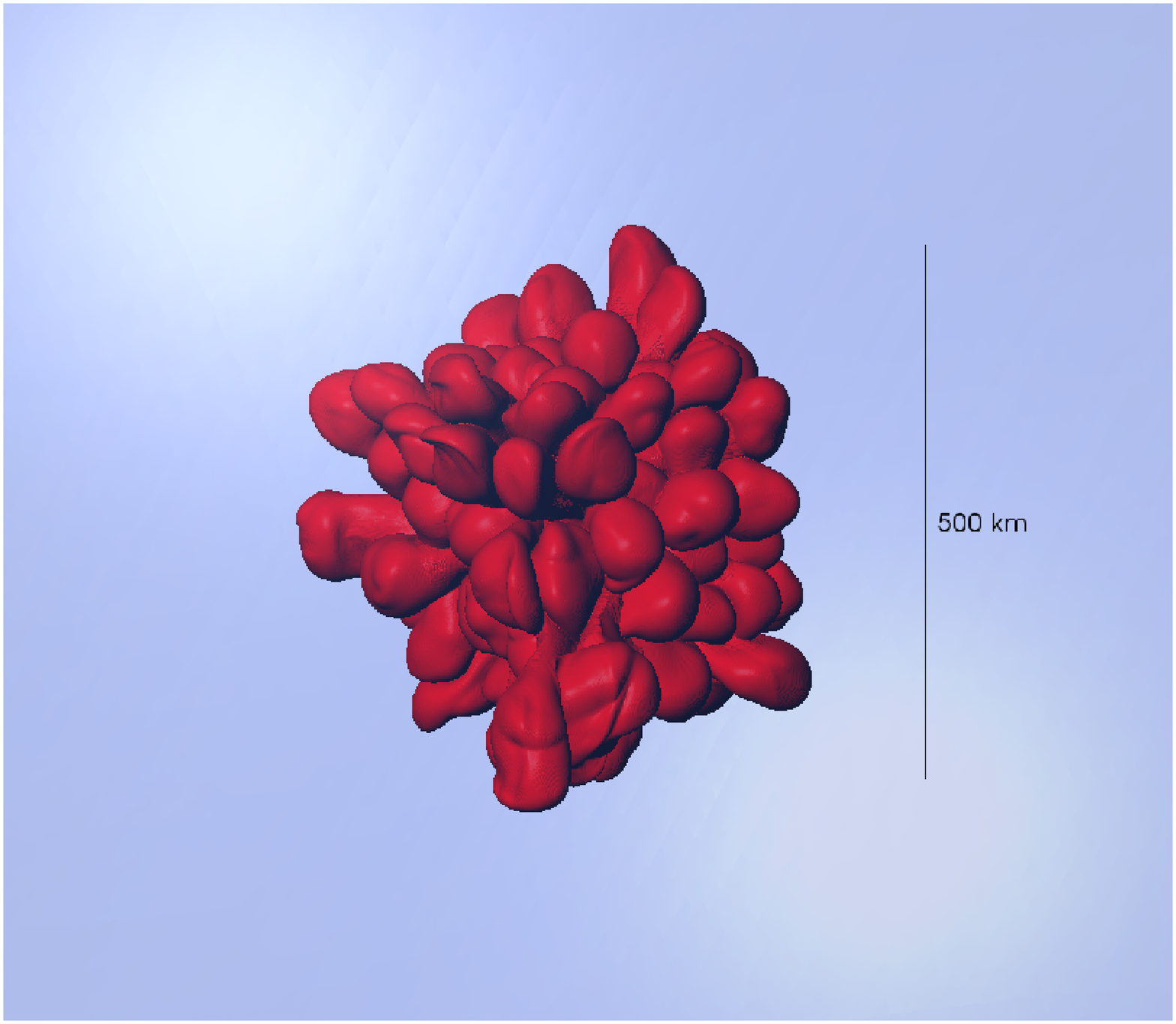}
\hfill
\includegraphics[width=0.4\textwidth,clip=true]{\figurepath 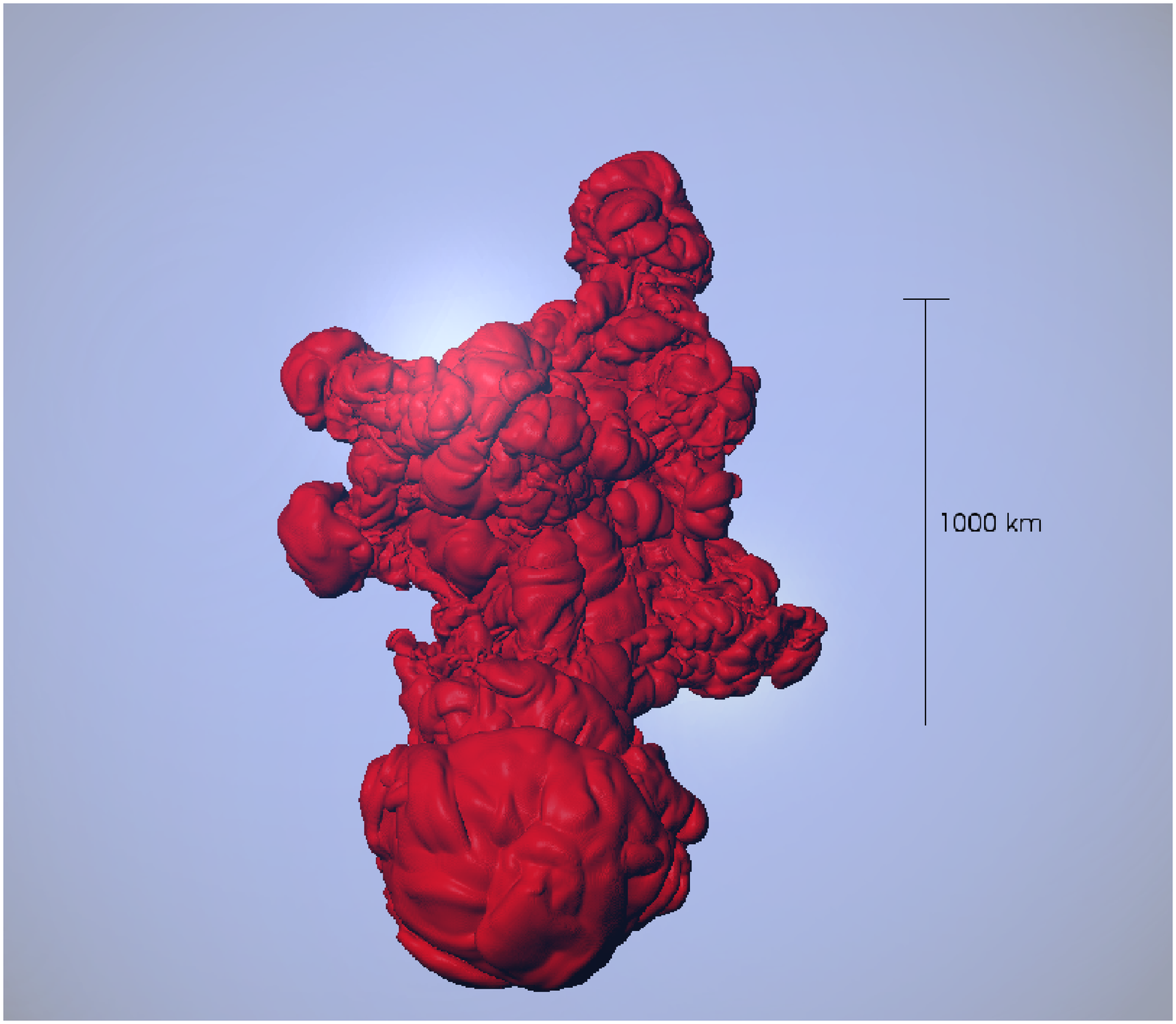}
\includegraphics[width=0.4\textwidth,clip=true]{\figurepath 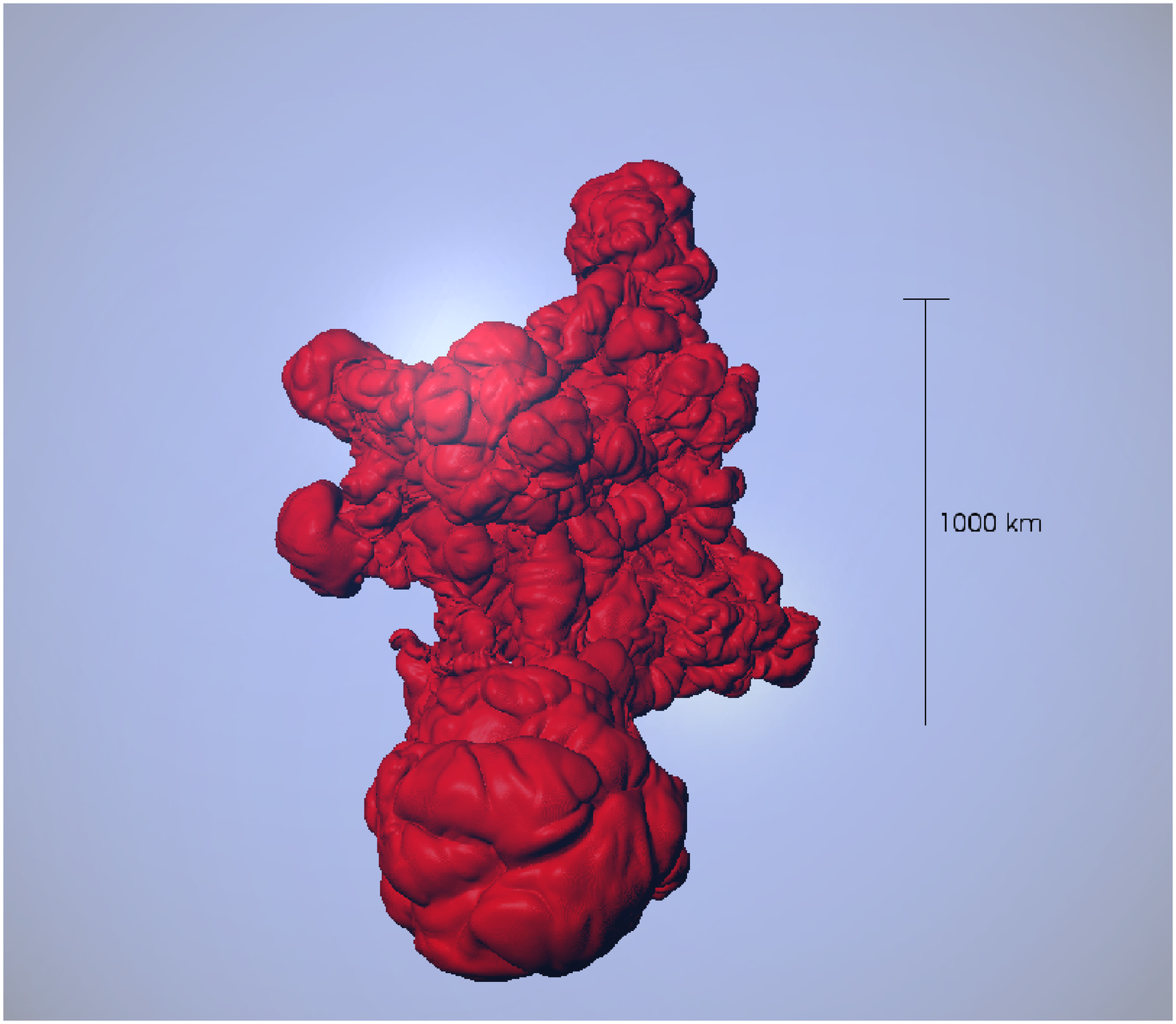}
\hfill
\includegraphics[width=0.4\textwidth,clip=true]{\figurepath 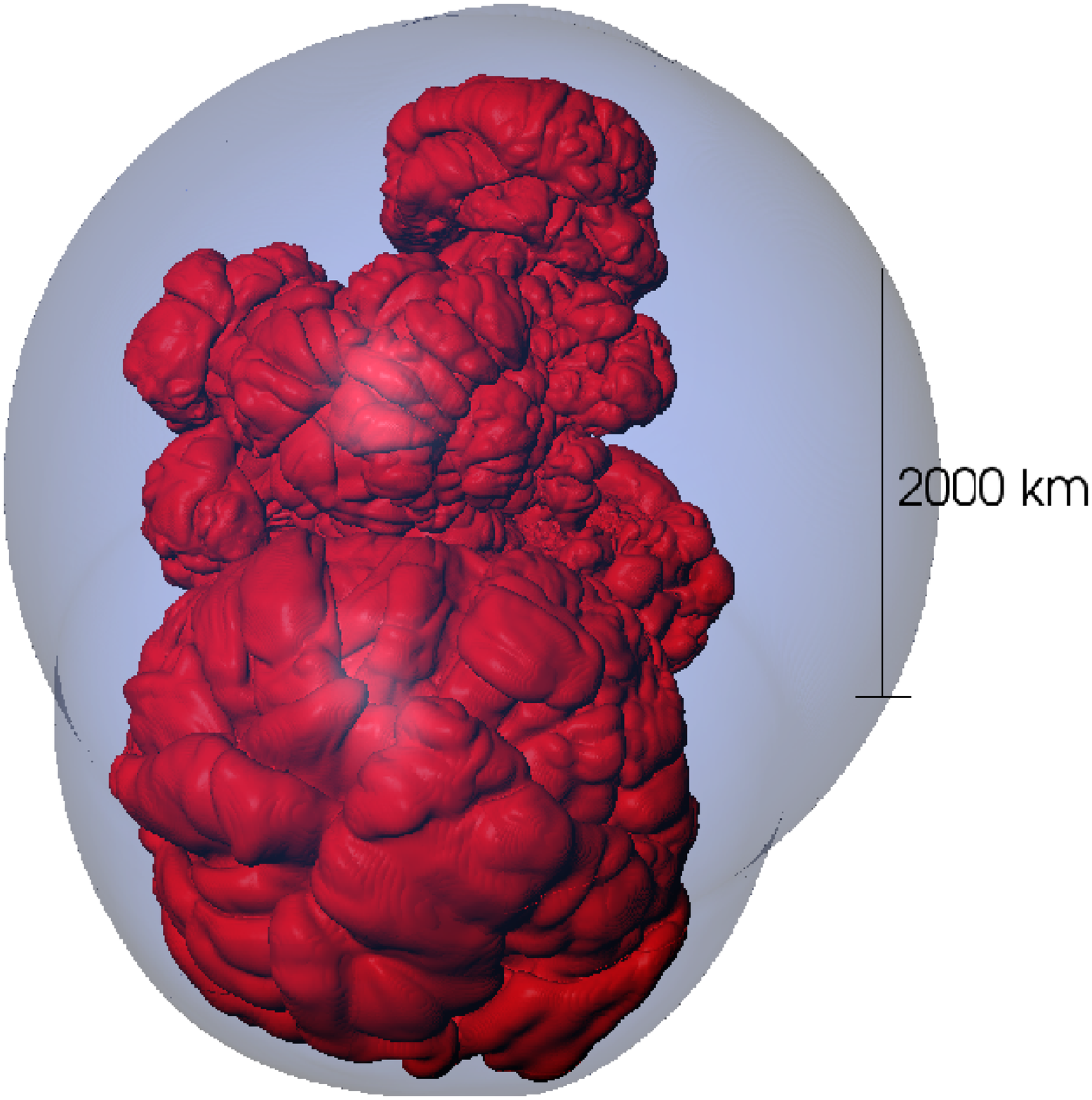}
\includegraphics[width=0.4\textwidth,clip=true]{\figurepath 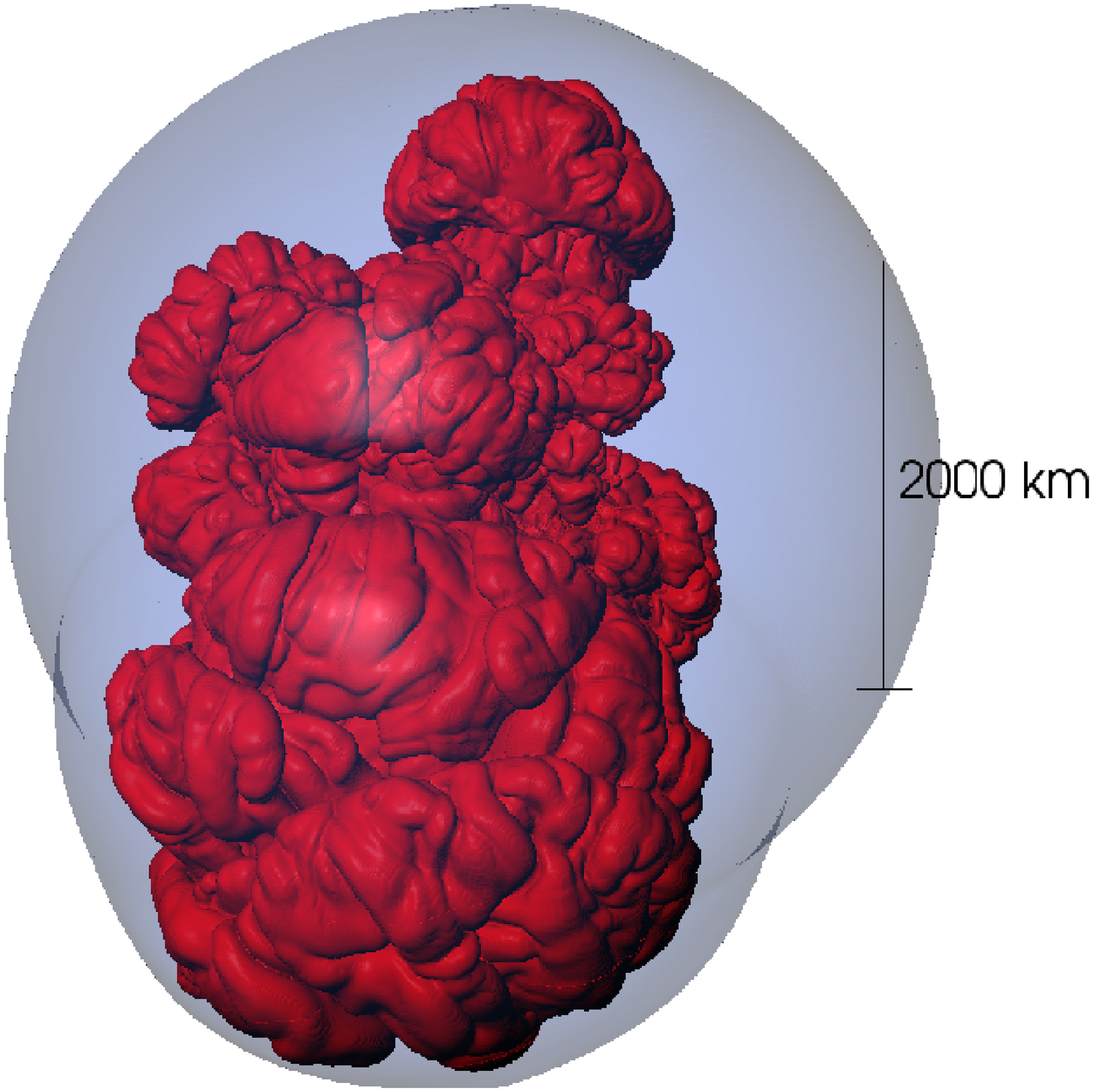}
\hfill
\caption[Similar to \Figure{f6}, but for B series]{Similar plots to
  \Figure{f6}, but for the B series of models.  The left column is
  model B100 and the right column is model BT; snapshots are (from top
  to bottom) at 0.3 s, 0.5 s, and 0.7 s.
\lFig{f8}}
\end{figure*}

\clearpage
\begin{figure}
\centering
\includegraphics[width=0.8\textwidth]{\figurepath 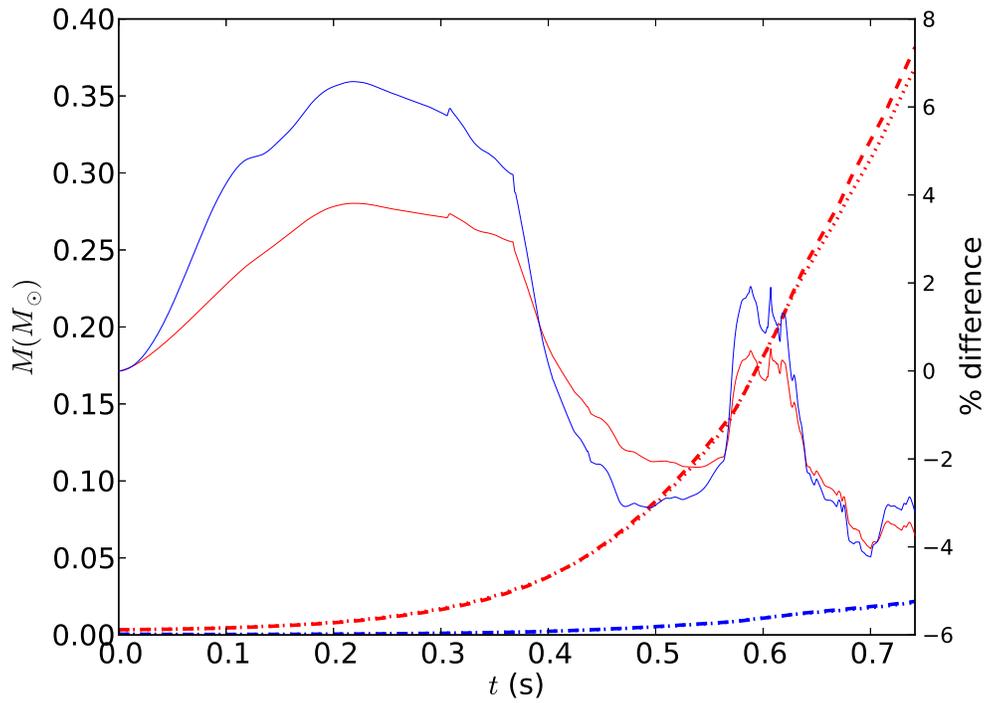}
\caption[B series mass evolution]{Total mass production of iron group
  elements (red) and IME (blue) for models B100 (dotted) and BT
  (dashed).  The thin lines show the percent difference between the
  B100 and BT models for each element group; a positive (negative)
  percent difference indicates that model B100 has produced more
  (less) material.  \lFig{f9}}
\end{figure}

\clearpage
\begin{figure}
\centering
\includegraphics[width=0.8\textwidth]{\figurepath 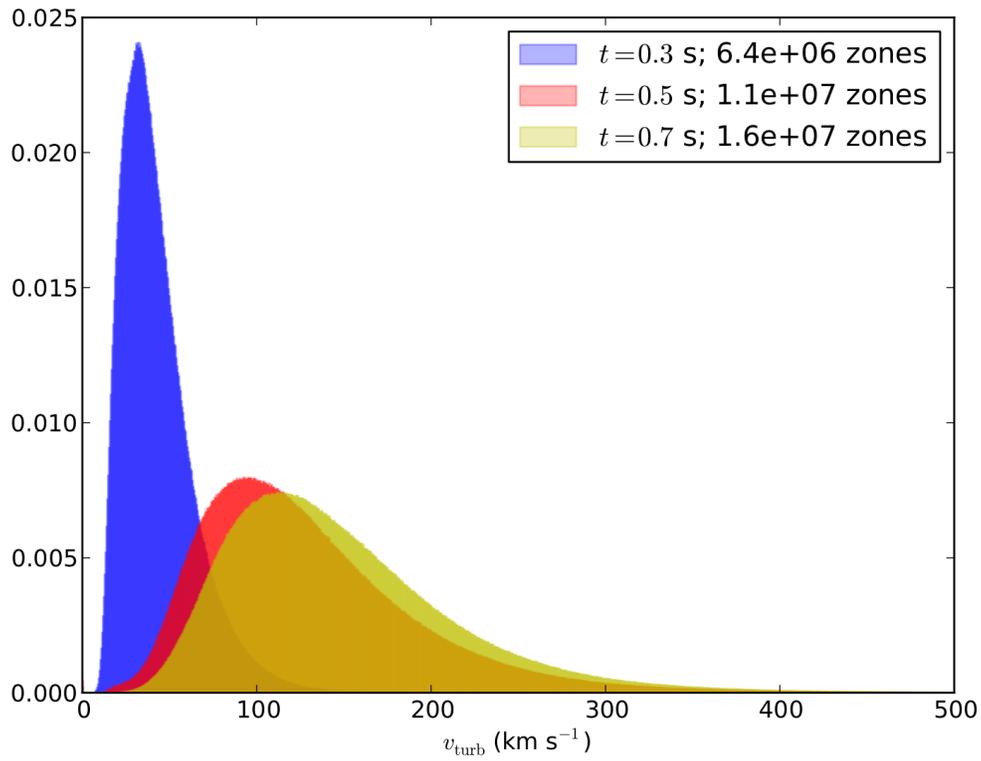}
\caption[PDF of turbulent velocity]{PDF of turbulent velocity inside
  the flame for model BT at various times.  Early on, the majority of
  the zones within the flame experience a turbulent intensity that is
  slower than the laminar speed, and therefore the flame propagates
  laminarly.  As the system evolves, the distribution widens and peaks
  towards larger flame speeds.\lFig{f10}}
\end{figure}

\clearpage
\begin{figure}
\centering
\includegraphics[angle=90,width=.62\textwidth,clip=true]{\figurepath 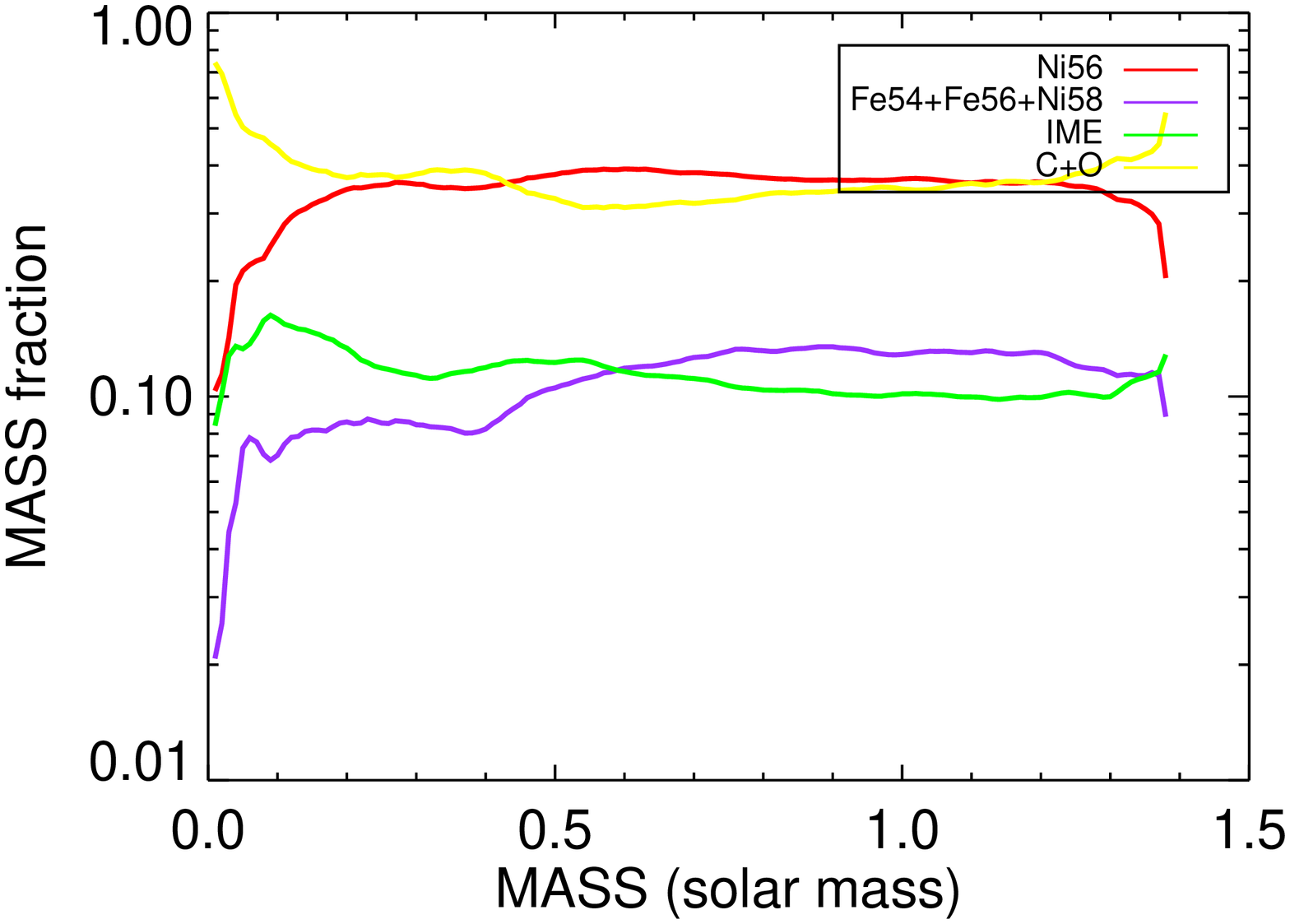}
\includegraphics[angle=90,width=.62\textwidth,clip=true]{\figurepath 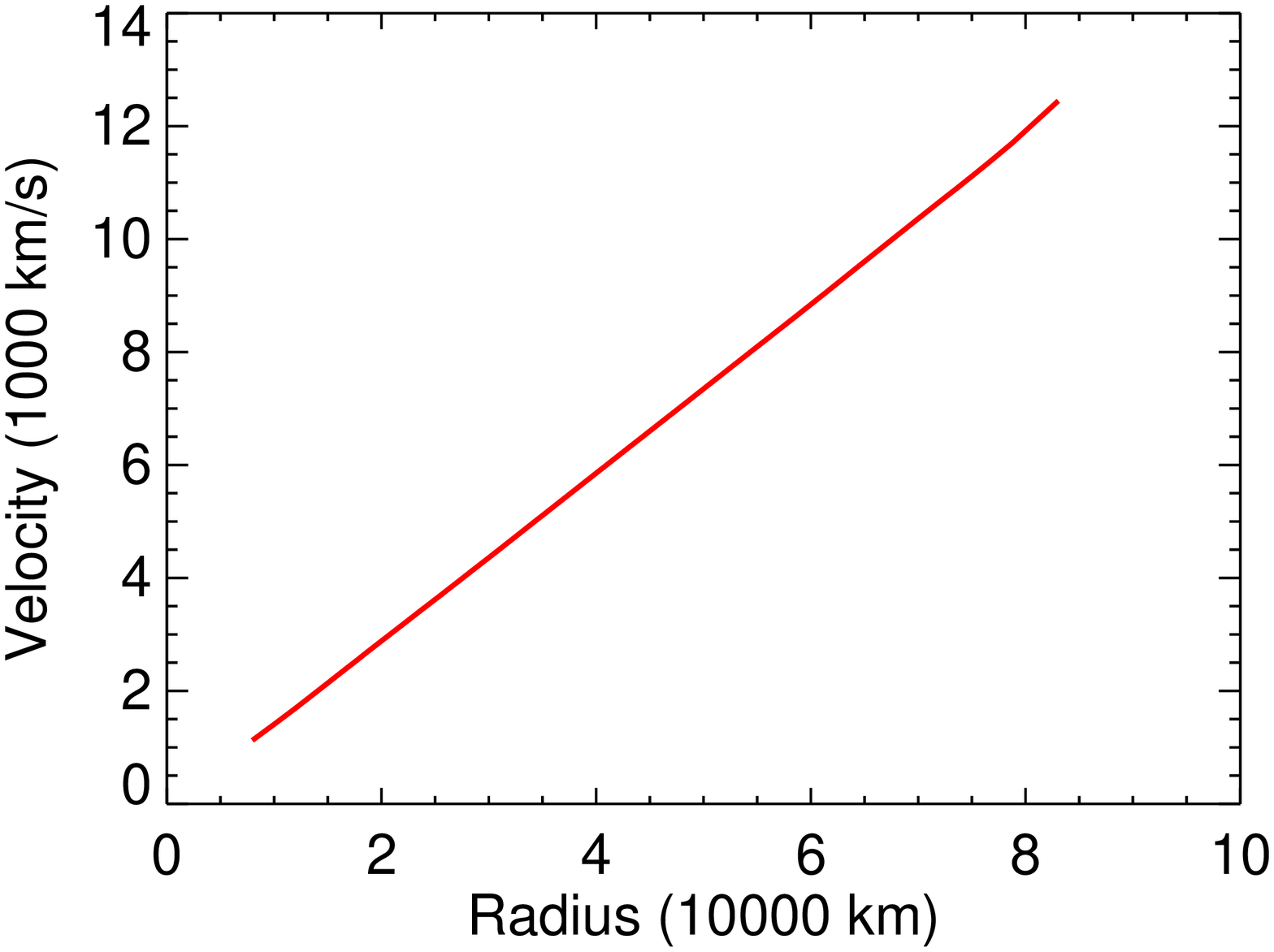}
\caption[Mass fraction of $\Nifs$, other iron group elements, IME, C+O] 
        {Mass fraction of $\Nifs$, other iron group elements, IME and 
         unburned C+O at 7.5 s for model A100. The plot of
         velocity versus radius shows the supernova ejecta is in homologous 
         expansion.
  \lFig{f11}}
\end{figure}
		    
{}

\clearpage


\clearpage

\begin{thebibliography}{}

\bibitem[Almgren et al.(2010)]{Alm10}
Almgren, A.~S., Beckner, V.~E., Bell, J.~B., et al.\ 2010, \apj, 715, 1221

\bibitem[Benetti et al.(2005)]{Ben05}
Benetti, S., et al. 2005, \apj, 623, 1011

\bibitem[Branch et al.(1985)]{Bra85} 
Branch, D., Doggett, J.~B., Nomoto, K., \& Thielemann, F.~K.\ 1985,
\apj, 294, 619

\bibitem[Branch \& Khokhlov(1995)]{Bra95}
Branch D., \& Khokhlov, A. M. 1995, Phys. Rep., 256, 53

\bibitem[Branch et al.(2003)]{Bra03}
Branch, D., et al. 2003, \aj, 126, 1489

\bibitem[Ciaraldi-Schoolmann et al.(2009)]{Cia09}
Ciaraldi-Schoolmann, F., Schmidt, W., Niemeyer, J.~C., R\"opke, F.~K., \&
Hillebrandt, W.\ 2009, \apj, 696, 1491

\bibitem[Damk\"ohler(1940)]{Dam40}
Damk\"ohler, G. 1940, Z. Elektrchem, 46, 601

\bibitem[Davies \& Taylor(1950)]{Dav50}
Davies, R. M., \& Taylor, G. 1950, Proc. R. Soc. London A, 200, 375

\bibitem[Filippenko(1997)]{Fil97}
Filippenko, A. V. 1997, \araa, 35, 309

\bibitem[Gamezo et al.(2003)]{Gam03}
Gamezo, V.~N., Khokhlov, A.~M., Oran, E.~S., Chtchelkanova, A.~Y., \&
Rosenberg, R.~O.\ 2003, Science, 299, 77

\bibitem[Gamezo et al.(2005)]{Gam05}
Gamezo, V.~N., Khokhlov, A.~M., \& Oran, E.~S.\ 2005, \apj, 623, 337

\bibitem[Garcia-Senz \& Woosley(1995)]{Gar95}
Garcia-Senz, D., \& Woosley, S.~E.\ 1995, \apj, 454, 895

\bibitem[Garc{\'i}a-Senz \& Bravo(2005)]{Gar05}
Garc{\'i}a-Senz, D., \& Bravo, E. 2005, \aap, 430, 585

\bibitem[Gilfanov \& Bogd{\'a}n(2010)]{Gil10} 
Gilfanov, M., \& Bogd{\'a}n, {\'A}.\ 2010, \nat, 463, 924

\bibitem[Hachisu et al.(2010)]{Hac10} 
Hachisu, I., Kato, M., \& Nomoto, K.\ 2010, \apjl, 724, L212

\bibitem[Hartmann, Woosley \& El Eid(1985)]{Har85}
Hartmann, D., Woosley, S.~E., \& El Eid, M.~F.\ 1985, \apj, 297, 837

\bibitem[Heger et al.(2001)]{Heg01}
Heger, A., Woosley, S.~E., Martinez-Pinedo, G., \& Langanke,
K.\ 2001, \apj, 560, 307

\bibitem[Hillebrandt \& Niemeyer(2000)]{Hil00} 
Hillebrandt, W., \& Niemeyer J. 2000, ARAA, 38, 191b

\bibitem[H\"oflich \& Khokhlov(1996)]{Hoe96} 
H\"oflich, P., \& Khokhlov, A.\ 1996, \apj, 457, 500

\bibitem[H{\"o}flich \& Stein(2002)]{Hoe02} 
H{\"o}flich, P., \& Stein, J.\ 2002, \apj, 568, 779

\bibitem[Howell et al.(2006)]{How06} 
Howell, D.~A., Sullivan, M., Nugent, P.~E., et al.\ 2006, \nat, 443,
308

\bibitem[Hoyle \& Fowler(1960)]{Hoy60}
Hoyle, F., \& Fowler, W.~A.\ 1960, \apj, 132, 565

\bibitem[Jackson et al.(2010)]{Jac10}
Jackson, A.~P., Calder, A.~C., Townsley, D.~M., Chamulak, D.~A., Brown, E.~F.
\& Timmes, F.~X.\ 2010, \apj, 720, 99

\bibitem[Kasen et al.(2009)]{Kas09}
Kasen, D.,  R\"opke, F., \& Woosley, S. E. 2009, \nat, 460, 869

\bibitem[Khokhlov(1995)]{Kho95}
Khokhlov, A. M.\ 1995, \apj, 449, 695

\bibitem[Khokhlov(1996)]{Kho96}
Khokhlov, A. M.\ 1996, Combust. Flame, 105, 28

\bibitem[Khokhlov et al.(1997)]{Kho97} 
Khokhlov, A.~M., Oran, E.~S., \& Wheeler, J.~C.\ 1997, \apj, 478, 678

\bibitem[Kromer et al.(2013)]{Kro13}
Kromer, M., Fink, M., Stanishev, V., Taubenberger, S., Ciaraldi-Schoolman, F.,
Pakmor, R., R{\"o}pke, F.~K., Ruiter, A.~J., Seitenzahl, I.~R., Sim, S.~A.,
Blanc, G., Elias-Rosa, N., \& Hillebrandt, W.\ 2013, \mnras, 429, 2287

\bibitem[Krueger et al.(2012)]{Kru12}
Krueger, B.~K., Jackson, A.~P., Calder, A.~C., Townsley, D.~M., Brown, E.~F.
\& Timmes, F.~X.\ 2012, \apj, 757, 175

\bibitem[Kuhlen, Woosley \& Glatzmaier(2006)]{Kuh06}
Kuhlen, M., Woosley, S. E., \& Glatzmaier, G. A. 2006, \apj, 640, 407

\bibitem[Landau \& Lifshitz(1959)]{Lan59}
Landau, L., \& Lifshitz, F. M. 1959, Course in Theoretical Physics,
Vol. 6.  Fluid Mechanics (Oxford:Pergammon)

\bibitem[Langanke \& Martinez-Pinedo(2000)]{Lan00}
Langanke, K. \& Martinez-Pinedo, G. 2000, Nucl. Phys. A, 673, 481

\bibitem[Marion et al.(2003)]{Mar03}
Marion, G. H., H\"oflich, P., Vacca, W. D. ,\& Wheeler,
J. C. 2003, \apj, 591, 316

\bibitem[Mazzali et al.(2007)]{Maz07}
Mazzali, P.~A., R{\"o}pke, F.~K., Benetti, S., \& Hillebrandt, W.\ 2007, Science, 315, 825

\bibitem[Niemeyer \& Woosley(1997)]{Nie97}
Niemeyer, J. C., \& Woosley, S. E. 1997, \apj, 475, 740

\bibitem[Nomoto, Thielemann \& Yokoi(1984)]{Nom84}
Nomoto, K., Thielemann, F.~K., \& Yokoi, K.\ 1984, \apj, 286, 644

\bibitem[Nonaka et al.(2012)]{Non12} 
Nonaka, A., Aspden, A.~J., Zingale, M., et al.\ 2012, \apj, 745, 73

\bibitem[Osher \& Sethian(1988)]{Osh88}
Osher, S., \& Sethian, J. A. 1988, J. Comp. Phys., 79, 12

\bibitem[Pakmor et al.(2011)]{Pak11} 
Pakmor, R., Hachinger, S., R{\"o}pke, F.~K., \& Hillebrandt, W.\ 2011,
\aap, 528, A117

\bibitem[Parrent et al.(2011)]{Par11}
Parrent, J. T., et al. 2011, \apj, 732, 30

\bibitem[Perets et al.(2010)]{Per10}
Perets, H.~B., Gal-Yam, A., Mazzali, P.~A., et al.\ 2010, \nat, 465,
322

\bibitem[Plewa et al.(2004)]{Ple04} 
Plewa, T., Calder, A.~C., \& Lamb, D.~Q.\ 2004, \apjl, 612, L37

\bibitem[Reinecke et al.(1999)]{Rei99}
Reinecke, M., Hillebrandt, W., Niemeyer, J.~C., Klein, R., \& Gr{\"o}bl, A.\ 1999, \aap, 347, 724

\bibitem[Reinecke, Hillebrandt \& Niemeyer(2002)]{Rei02}
Reinecke, M., Hillebrandt, W., \& Niemeyer, J.~C.\ 2002, \aap, 391, 1167

\bibitem[R\"{o}pke \& Hillebrandt(2005)]{Rop05}
R\"{o}pke, F. K., \& Hillebrandt, W. 2005, \aap, 431, 635

\bibitem[R{\"o}pke et al.(2006)]{Rop06} 
R{\"o}pke, F.~K., Hillebrandt, W., Niemeyer, J.~C., \& Woosley,
S.~E.\ 2006, \aap, 448, 1

\bibitem[R{\"o}pke(2007)]{Rop07}
R{\"o}pke, F.~K.\  2007, \apj, 668, 1103

\bibitem[R{\"o}pke et al.(2007a)]{Rop07a}
R{\"o}pke, F.~K., Hillebrandt, W., Schmidt, W., et al.\ 2007a, \apj, 668, 1132

\bibitem[R{\"o}pke et al.(2007b)]{Rop07b} 
R{\"o}pke, F.~K., Woosley, S.~E., \& Hillebrandt, W.\ 2007b, \apj, 660,
1344

\bibitem[Schmidt et al.(2006a)]{Sch06a} 
Schmidt, W., Niemeyer, J.~C., \& Hillebrandt, W.\ 2006a, \aap, 450, 265

\bibitem[Schmidt et al.(2006b)]{Sch06b} 
Schmidt, W., Niemeyer, J.~C., Hillebrandt, W., R\"opke, F.~K.\
2006b, \aap, 450, 283

\bibitem[Schmidt \& Niemeyer(2006)]{Sch06}
Schmidt, W., \& Niemeyer, J.~C.\ 2006, \aap, 446, 627

\bibitem[Sharp(1984)]{Sha84}
Sharp, D. H., 1984, Physica D:Nonlinear Phenomena, 12, 3

\bibitem[Stritzinger et al.(2006)]{Str06} 
Stritzinger, M., Leibundgut, B., Walch, S., \& Contardo, G.\ 2006,
\aap, 450, 241

\bibitem[Timmes \& Woosley(1992)] {Tim92}
Timmes, F.~X., \& Woosley, S.~E.\ 1992, \apj, 396, 649

\bibitem[Timmes \& Swesty(2000)]{Tim00} 
Timmes, F.~X., \& Swesty, F.~D., 2000, \apjs, 126, 501

\bibitem[Townsley et al.(2007)]{Tow07} 
Townsley, D.~M., Calder, A.~C., Asida, S.~M., et al.\ 2007, \apj, 668,
1118

\bibitem[Voss \& Nelemans(2008)]{Vos08} 
Voss, R., \& Nelemans, G.\ 2008, \nat, 451, 802


\bibitem[Weaver, Woosley \& Zimmerman(1978)]{Wea78} 
Weaver, T.~A., Woosley, S.~E., \& Zimmerman, G.~B.\ 1978, \apj, 225, 1021

\bibitem[Woosley et al.(2002)]{Woo02}
Woosley, S.~E., Heger, A., \& Weaver, T.~A.\ 2002, Reviews of Modern
Physics, 74, 1015

\bibitem[Woosley et al.(2007)]{Woo07a} 
Woosley, S.~E., Kasen, D., Blinnikov, S., \& Sorokina, E.\ 2007, \apj,
662, 487

\bibitem[Woosley(2007)]{Woo07}
Woosley, S.~E.\ 2007, \apj, 668, 1109

\bibitem[Woosley et al.(2009)]{Woo09}
Woosley, S.~E., Kerstein, A.~R., Sankaran, V., Aspden, A.~J., \&
R\"opke, F.~K.\ 2009, \apj, 704, 255

\bibitem[Woosley \& Kasen(2011)]{Woo11a} 
Woosley, S.~E., \& Kasen, D.\ 2011, \apj, 734, 38

\bibitem[Woosley et al.(2011)]{Woo11b}
Woosley, S. E., Kerstein, A. R., \& Aspden, A. J. 2011, \apj, 734, 37

\bibitem[Yoon et al.(2007)]{Yoo07} 
Yoon, S.-C., Podsiadlowski, P., \& Rosswog, S.\ 2007, \mnras, 380, 933

\bibitem[Zingale et al.(2005)]{Zin05}
Zingale, M., Woosley, S.~E., Rendleman, C.~A., Day, M.~S., \& Bell, J.~B.\ 
2005, \apj, 632, 1021

\bibitem[Zingale \& Dursi(2007)]{Zin07} 
Zingale, M., \& Dursi, L.~J.\ 2007, \apj, 656, 333

\bibitem[Zingale et al.(2009)]{Zin09}
Zingale, M., Almgren, A. S., Bell, J. B., Nonaka, A., \& Woosley, S. E. 2009, \apj, 704, 196

\bibitem[Zingale et al.(2011)]{Zin11}
Zingale, M., Nonaka, A., Almgren, A. S., et al.  2011, \apj, 740, 8

\end{thebibliography}
\end{document}